\documentclass[journal,article,submit,moreauthors,pdftex,10pt,a4paper]{mdpi} 
\firstpage{1} 
\articlenumber{x}
\doinum{10.3390/------}
\pubvolume{xx}
\pubyear{2017}
\copyrightyear{2017}
\externaleditor{Academic Editor: name}
\history{Received: date; Accepted: date; Published: date}

\pdfoutput=1

\Title{Femtoscopy with identified hadrons in pp, pPb, \\ and PbPb collisions in CMS}

\Author{Ferenc Sikl\'er $^{1}$\orcidA{} for the CMS Collaboration}

\AuthorNames{Ferenc Sikl\'er}

\address{%
$^{1}$ \quad Wigner RCP, Budapest; sikler.ferenc@wigner.mta.hu}

\newcommand{\pt}{p_\mathrm{T}}
\newcommand{\kt}{k_\mathrm{T}}

\newcommand{\qi}{q_\mathrm{inv}}
\newcommand{\ql}{q_\mathrm{l}}
\newcommand{\qt}{q_\mathrm{t}}
\newcommand{\qo}{q_\mathrm{o}}
\newcommand{\qs}{q_\mathrm{s}}

\newcommand{\Rp}{R_\mathrm{param}}

\newcommand{\ldg}{Lines are drawn to guide the eye.}

\newcommand{\GeVc}{GeV/$c$}

\renewcommand{\vec}[1]{{\bf{#1}}}

\newcommand{\sqs}{\sqrt{s}}
\newcommand{\ssNN}{\sqrt{s_{_\mathrm{NN}}}}
\newcommand{\Nt}{N_\mathrm{tracks}}
\newcommand{\Nr}{N_\mathrm{rec}}

\abstract{
Short range correlations of identified charged hadrons in pp ($\sqrt{s} =$ 0.9,
2.76, and 7~TeV), pPb ($\sqrt{s_{_\mathrm{NN}}} =$ 5.02~TeV), and peripheral
PbPb collisions ($\sqrt{s_{_\mathrm{NN}}} =$ 2.76~TeV) are studied with the CMS
detector at the LHC. Charged pions, kaons, and protons at low momentum and in
laboratory pseudorapidity $|\eta| < 1$ are identified via their energy loss in
the silicon tracker. The two-particle correlation functions show effects of
quantum statistics, Coulomb interaction, and also indicate the role of
multi-body resonance decays and mini-jets. The characteristics of the one-,
two-, and three-dimensional correlation functions are studied as a function of
transverse pair momentum, $\kt$, and the charged-particle multiplicity of the
event.  The extracted radii are in the range 1-5 fm, reaching highest values
for very high multiplicity pPb, also for similar multiplicity PbPb collisions,
and decrease with increasing $\kt$. The dependence of radii on multiplicity and
$\kt$ largely factorizes and appears to be insensitive to the type of the
colliding system and center-of-mass energy.  }

\conferencetitle{BGL17: 10th Bolyai-Gauss-Lobachevsky conference}

\begin{document}

\section{Introduction}

Measurements of the correlation between hadrons emitted in high energy
collisions of nucleons and nuclei can be used to study the spatial extent
and shape of the created system. The characteristic radii, the homogeneity
lengths, of the particle emitting source can be extracted with reasonable
precision~\cite{Erazmus:1996ns}.
The topic of quantum correlations was well researched in the past by the CMS
Collaboration~\cite{Khachatryan:2010un,Khachatryan:2011hi} using unidentified
charged hadrons produced in $\sqs =$ 0.9, 2.36, and 7 TeV pp collisions. Those
studies only included one-dimensional fits ($\qi$) of the correlation function.
Our aim was to look for effects present in pp, pPb, and PbPb interactions using
the same analysis methods, producing results as a function of the transverse
pair momentum $\kt$ and of the fully corrected charged-particle multiplicity
$\Nt$ (in $|\eta|<$ 2.4) of the event.
In addition, not only charged pions, but also charged kaons are
studied. All details of the analysis are given in Ref.~\cite{CMS:2014mla}.

\section{Data analysis}

The analysis methods (event selection, reconstruction of charged particles in
the silicon tracker, finding interaction vertices, treatment of pile-up) are
identical to the ones used in the previous CMS papers on the spectra of
identified charged hadrons produced in $\sqs =$ 0.9, 2.76, and 7 TeV
pp~\cite{Chatrchyan:2012qb} and $\ssNN =$ 5.02 TeV pPb
collisions~\cite{Chatrchyan:2013eya}.
A detailed description of the CMS detector can be found
in Ref.~\cite{:2008zzk}.

For the present study 8.97, 9.62, and 6.20 M minimum bias events are used from
pp collisions at $\sqs =$ 0.9 TeV, 2.76 TeV, and 7 TeV, respectively, while
8.95 M minimum bias events are available from pPb collisions at $\ssNN =$
5.02 TeV.
The data samples are completed by 3.07 M peripheral (60--100\%) PbPb events,
where 100\% corresponds to fully peripheral, 0\% means fully central (head-on)
collision. The centrality percentages for PbPb are determined via measuring the
sum of the energies in the forward calorimeters.

The multiplicity of reconstructed tracks, $\Nr$, is obtained in the region
$|\eta|<$ 2.4. Over the range 0 $< \Nr <$ 240, the events were divided into 24
classes, a region that is well covered by the 60--100\% centrality PbPb
collisions.
To facilitate comparisons with models, the corresponding corrected charged
particle multiplicity $\Nt$ in the same acceptance of $|\eta|<$ 2.4 is also
determined.

\begin{figure}

 \begin{center}
  \includegraphics[width=0.49\textwidth]{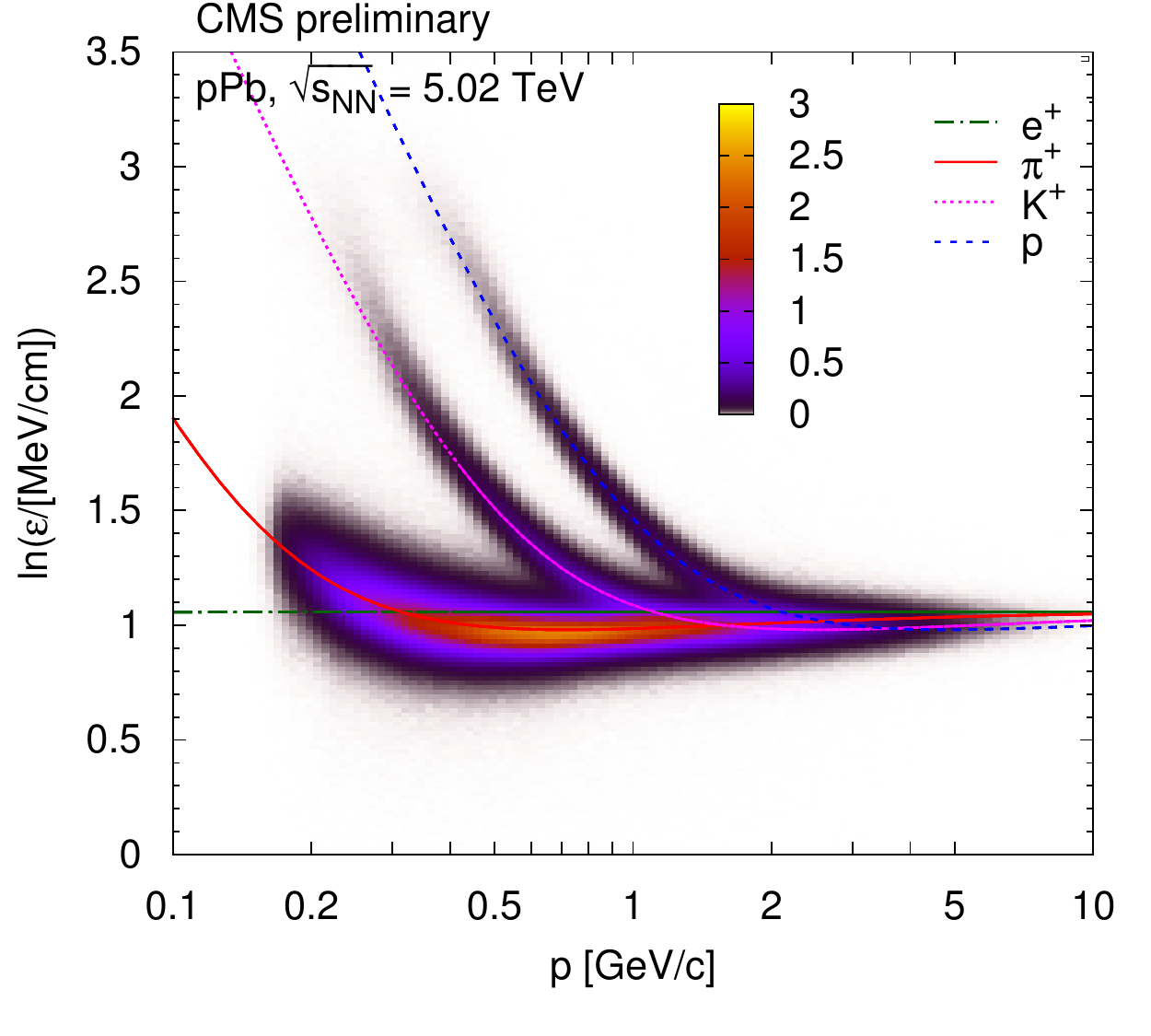}
  \includegraphics[width=0.49\textwidth]{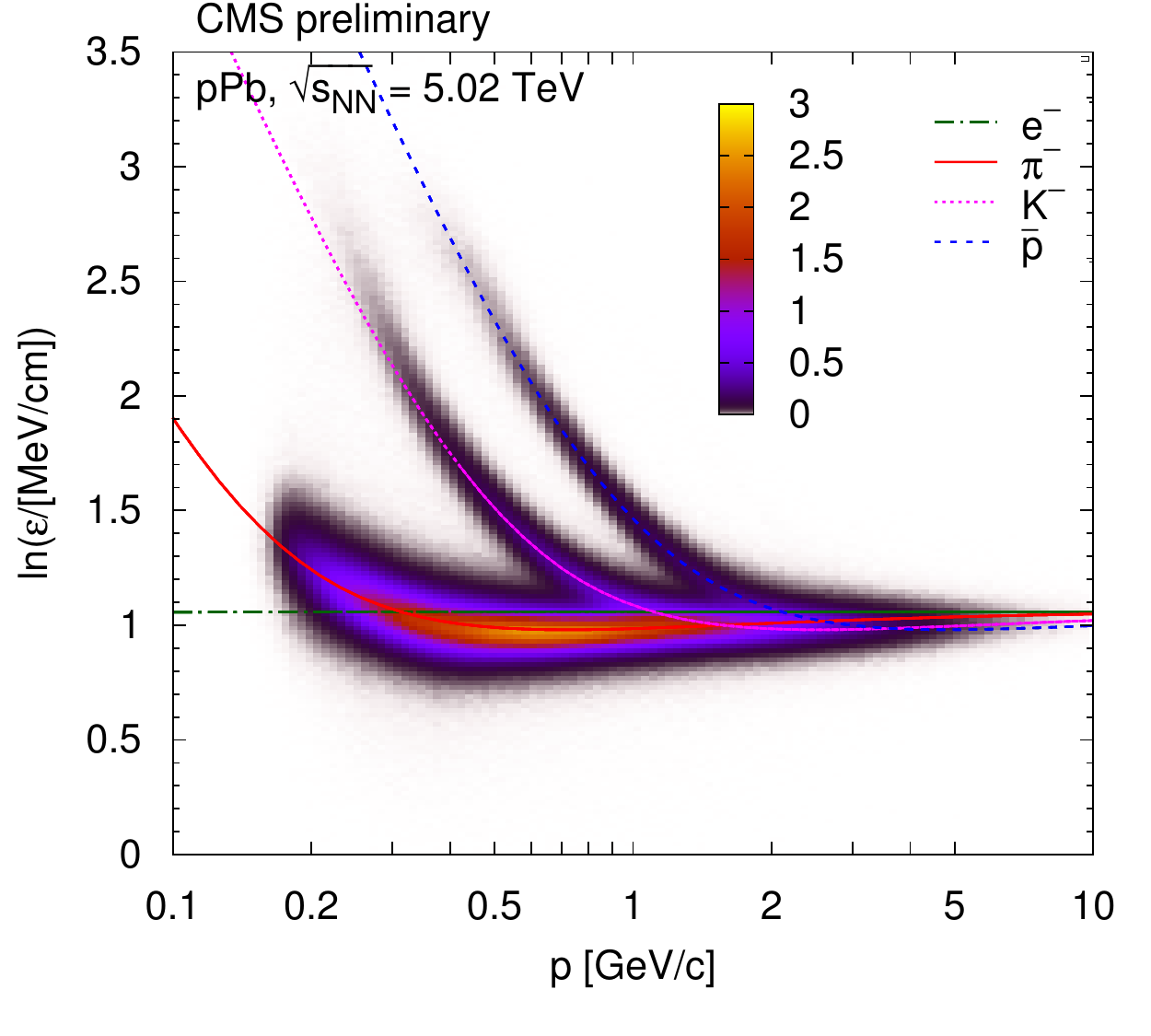}
 \end{center}

 \vspace{-0.2in}
 \caption{The distribution of $\ln\varepsilon$ as a function of total momentum
$p$, for positively (left) and negatively (right) charged particles, in case of
pPb collisions at $\ssNN =$ 5.02 TeV~\cite{CMS:2014mla}. Here $\varepsilon$ is
the most probable energy loss rate at a reference path length $l_0 =$
450~$\mu$m. The $z$ scale is shown in arbitrary units and is linear. The curves
show the expected $\ln\varepsilon$ for electrons, pions, kaons, and protons
(full theoretical calculation, Eq.~(30.11) in Ref.~\cite{Beringer:1900zz}).}

 \label{fig:elossHistos}

\end{figure}

The reconstruction of charged particles in CMS is bounded by the acceptance of
the tracker and by the decreasing tracking efficiency at low momentum.
Particle-by-particle identification using specific ionization is possible in
the momentum range $p <$ 0.15~\GeVc\ for electrons, $p <$ 1.15~\GeVc\ for pions
and kaons, and $p <$ 2.00~\GeVc\ for protons (Fig.~\ref{fig:elossHistos}).
In view of the $(\eta,\pt)$ regions where pions, kaons, and protons can all be
identified, only particles in the band $-$1 $< \eta <$ 1 (in the laboratory
frame) were used for this measurement.
In this analysis a very high purity ($>$ 99.5\%) particle identification is
required, ensuring that less than 1\% of the examined particle pairs would be
fake.

\subsection{Correlations}

The pair distributions are binned in the number of reconstructed charged
particles $\Nr$ of the event, in the transverse pair momentum $\kt =
|\vec{p_{T,1}} + \vec{p_{T,2}}|/2$, and also in the relative momentum
($\vec{q}$) variables in the longitudinally co-moving system of the
pair. One-dimensional ($\qi = |\vec{q}|$), two-dimensional $(\ql,\qt)$, and
three-dimensional $(\ql,\qo,\qs)$ analyses are performed. Here $\qo$ is the
component of $\vec{\qt}$ parallel to $\vec{\kt}$, $\qs$ is the component of
$\vec{\qt}$ perpendicular to $\vec{\kt}$.

The construction of the $\vec{q}$ distribution for the ``signal'' pairs is
straightforward: all valid particle pairs from the same event are taken and the
corresponding histograms are filled.
There are several choices for the construction of the background. We considered
the following three prescriptions:

\begin{itemize}
 \item particles from the actual event are paired with particles from some
       given number of, in our case 25, preceding events (``event mixing'');
       only events belonging to the same multiplicity ($\Nr$) class are mixed;
 \item particles from the actual event are paired, the
       laboratory momentum vector of
       the second particle is rotated
       around the beam axis by 90 degrees (``rotated'');
 \item particles from the actual event are paired, but the
       laboratory momentum vector of the second particle is negated
       (``mirrored'').
\end{itemize}

\noindent
Based on the goodness-of-fit distributions the
event mixing prescription was used while the rotated and mirrored versions,
which give worse or much worse $\chi^2$/ndf values, were employed in the
estimation of the systematic uncertainty.

The {measured} two-particle correlation function $C_2(\vec{q})$ is the
ratio of signal and background distributions

\begin{equation}
 C_2(\vec{q}) = \frac{N_\mathrm{signal}(\vec{q})}
                     {N_\mathrm{bckgnd}(\vec{q})},
\end{equation}

\noindent where the background is normalized such that it has the same integral
as the signal distribution.
The quantum correlation function $C_\mathrm{BE}$, part of $C_2$, is the Fourier
transform of the source density distribution $f(\vec{r})$.  There are several
possible functional forms that are commonly used to fit 
$C_\mathrm{BE}$ present in the data: Gaussian ($1 + \lambda \exp\left[-
(q R)^2/(\hbar c)^2 \right]$) and exponential parametrizations ($1 + \lambda
\exp\left[- (|q| R)/(\hbar c)\right]$), and a mixture of those in higher
dimensions. (The denominator $\hbar c =$ 0.197~GeV~fm is usually omitted from
the formulas, we will also do that in the following.)
Factorized forms are particularly popular, such as $\exp(- \ql^2 R_l^2 -\qo^2
R_o^2 - \qs^2 R_s^2)$ or $\exp(- \ql R_l -\qo R_o - \qs R_s)$ with some
theoretical motivation.
The fit parameters are usually interpreted as chaoticity $\lambda$, and
characteristic radii $R$, the homogeneity lengths, of the particle emitting
source.

As will be shown in Sec.~\ref{sec:results}, the exponential parametrization
does a very good job in describing all our data. It corresponds to the Cauchy (Lorentz) type source distribution $f(r) = R/(2 \pi^2 \left[r^2 +
(R/2)^2\right]^2)$.
Theoretical studies show that for the class of {stable distributions}, with
index of stability $0 < \alpha \le 2$, the Bose-Einstein correlation function
has a {stretched exponential} shape~\cite{Csorgo:2003uv,Csorgo:2004ch}.
The exponential correlation function implies $\alpha = 1$. (The Gaussian would
correspond to the special case of $\alpha = 2$.) The forms used for the
fits are
\begin{align}
 C_\mathrm{BE}(\qi)     
   &= 1 + \lambda \exp\left[- \qi R \right], \\
 C_\mathrm{BE}(\ql,\qt) 
   &= 1 + \lambda \exp\left[- \sqrt{(q_l R_l)^2 + (q_t R_t)^2}\right], \\
 C_\mathrm{BE}(\ql,\qo,\qs) 
   &= 1 + \lambda \exp\left[- \sqrt{(q_l R_l)^2 + (q_o R_o)^2 + (q_s R_s)^2}\right],
\end{align}

\noindent meaning that the system in multi-dimensions is an ellipsoid with
differing radii $R_l$, $R_t$, or $R_l$, $R_o$, and $R_s$.

\subsection{Coulomb interaction}

After the removal of the trivial phase space effects (ratio of signal and
background distributions), one of the most important source of correlations is
the mutual Coulomb interaction of the emitted charged particles.
The effect of the Coulomb interaction is taken into account by the factor $K$,
the squared average of the relative wave function $\Psi$, as $K(\qi) = \int
d^3\vec{r} \; f(\vec{r}) \; |\Psi(\vec{k},\vec{r})|^2$, where $f(\vec{r})$ is
the source intensity discussed above.
For pointlike source, $f(\vec{r}) = \delta(\vec{r})$, and we get the Gamow
factor $G(\eta) = |\Psi(0)|^2 = 2\pi\eta/[\exp(2\pi\eta) - 1]$, where $\eta =
\pm \alpha m/\qi$ is the Landau parameter, $\alpha$ is the fine-structure
constant, $m$ is the mass of the particle. The positive sign should be used for
repulsion, and the negative is for attraction.

For an extended source, a more elaborate treatment is
needed~\cite{PhysRevD.33.72}. The use of the Bowler-Sinyukov
formula~\cite{Bowler:1991vx,Sinyukov:1998fc} is popular.
Our data on unlike-sign correlation functions show that while the Gamow factor
might give a reasonable description of the Coulomb interaction for pions, it is
clearly not enough for kaons.
In the $q$ range studied in this analysis $\eta \ll 1$ applies. The absolute
square of confluent hypergeometric function of the first kind $F$, present in
$\Psi$, can be well approximated as $|F|^2 \approx 1 + 2\eta
\operatorname{Si}(x)$ where $\operatorname{Si}$ is the sine integral function.
Furthermore, for Cauchy type source functions the factor $K$ is nicely
described by the formula $K(\qi) = G(\eta) \left[1 + \pi\eta \qi R / (1.26 +
\qi R)\right]$. In the last step we substituted $\qi = 2k$. The factor $\pi$ in
the approximation comes from the fact that for large $kr$ arguments
$\operatorname{Si}(kr) \rightarrow \pi/2$.  Otherwise it is a simple but
faithful approximation of the result of a numerical calculation, with
deviations less than 0.5\%.

\begin{figure}

 \begin{center}
  \includegraphics[width=0.49\textwidth]{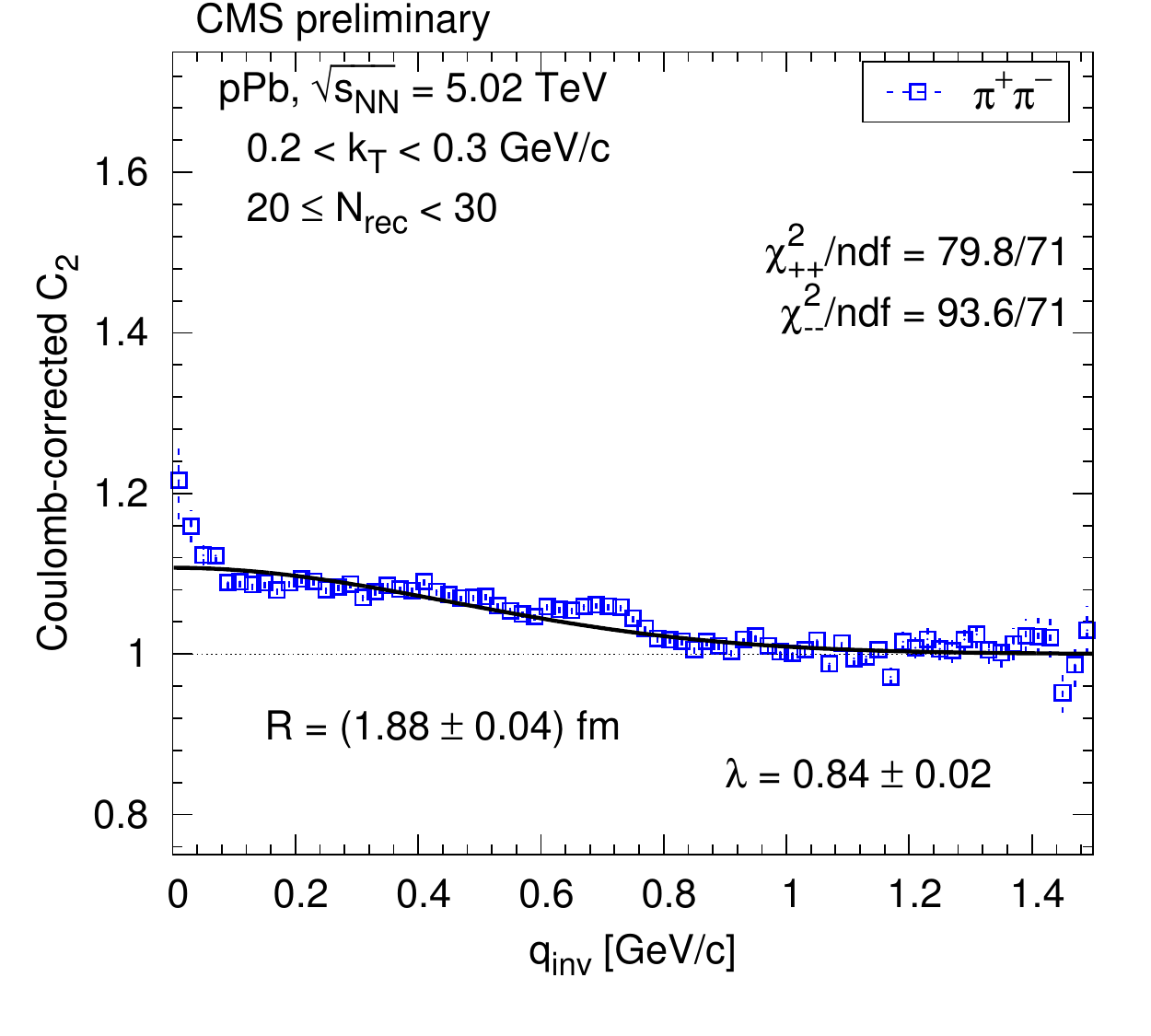}
  \includegraphics[width=0.49\textwidth]{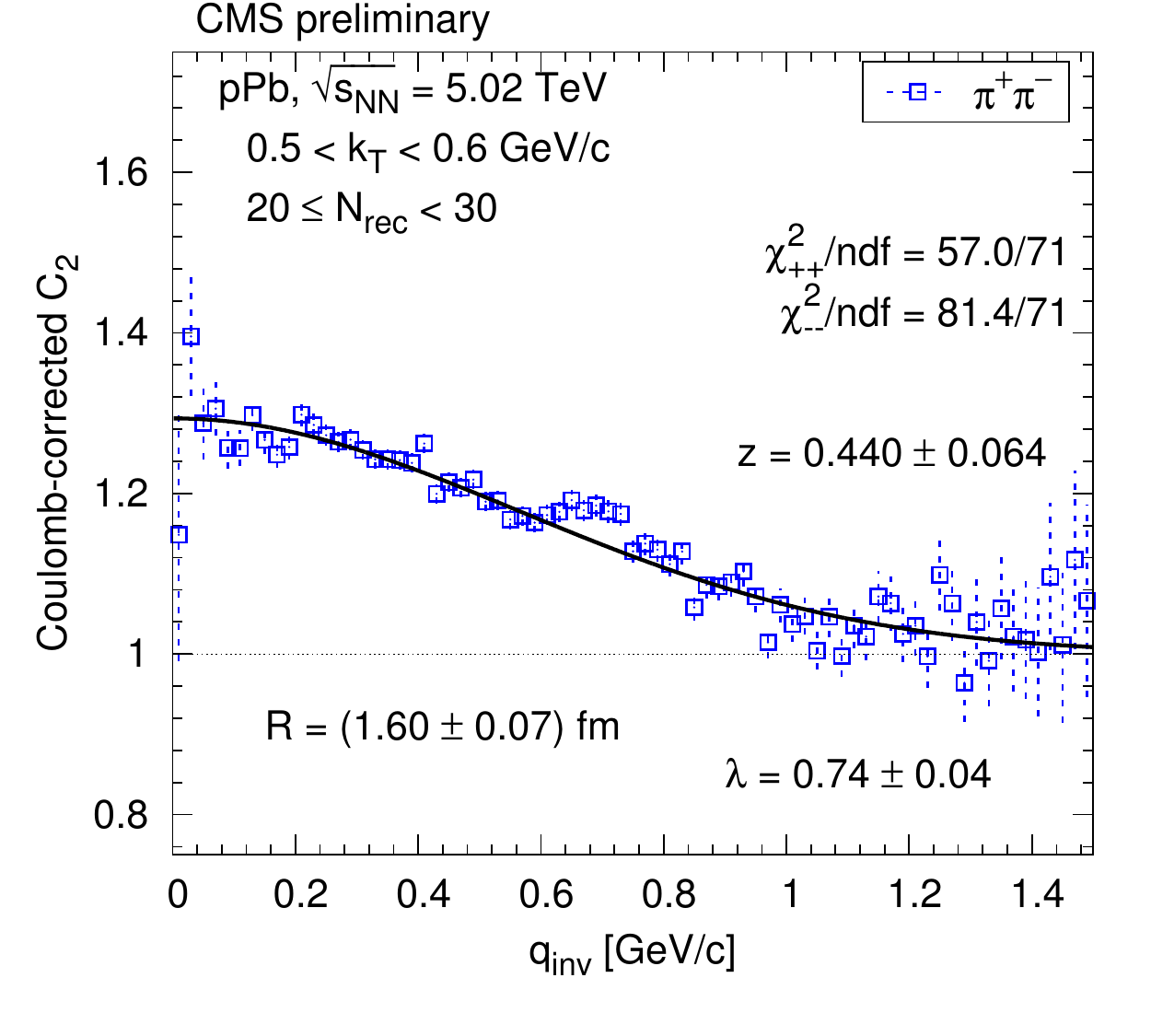}
 \end{center}

 \vspace{-0.2in}
 \caption{Contribution of clusters (mini-jets and multi-body decays of
resonances) to the measured Coulomb-corrected correlation function of
$\pi^+\pi^-$ (open squares) for some selected $\kt$ bins, 20 $\le \Nr <$ 30,
in case of pPb interactions at $\ssNN =$ 5.02 TeV~\cite{CMS:2014mla}. The solid
curves show the result of the Gaussian fit.}

 \label{fig:minijet_vs_Nrec_and_kt}

\end{figure}

\subsection{Clusters: mini-jets, multi-body decays of resonances}

The measured unlike-sign correlation functions show contributions from various
resonances.  The seen resonances include the $\mathrm{K^0_S}$, the
$\rho(770)$, the $\mathrm{f_0(980)}$, the $\mathrm{f_2(1270)}$
decaying to $\mathrm{\pi^+ \pi^-}$, and the $\phi(1020)$ decaying to
$\mathrm{K^+ K^-}$. Also, $\mathrm{e^+ e^-}$ pairs from $\gamma$ conversions,
when misidentified as pion pairs, can appear as a very low $\qi$ peak in the
$\pi^+\pi^-$ spectrum.
With increasing $\Nr$ values the effect of resonances diminishes, since their
contribution is quickly exceeded by the combinatorics of unrelated particles.

Nevertheless, the Coulomb-corrected unlike-sign correlation functions are not
always close to unity at low $\qi$, but show a Gaussian-like hump
(Fig.~\ref{fig:minijet_vs_Nrec_and_kt}). That structure has a varying amplitude
but a stable scale ($\sigma$ of the corresponding Gaussian) of about
0.4~\GeVc. This feature is often related to particles emitted inside low
momentum mini-jets, but can be also attributed to the effect of multi-body
decays of resonances. In the following we will refer to those possibilities as
fragmentation of clusters, or {cluster contribution}.
We have fitted the one-dimensional unlike-sign correlation functions with a
$(\Nr,\kt)$-dependent Gaussian parametrization~\cite{CMS:2014mla}.

\begin{figure}

 \begin{center}
  \includegraphics[width=0.49\textwidth]{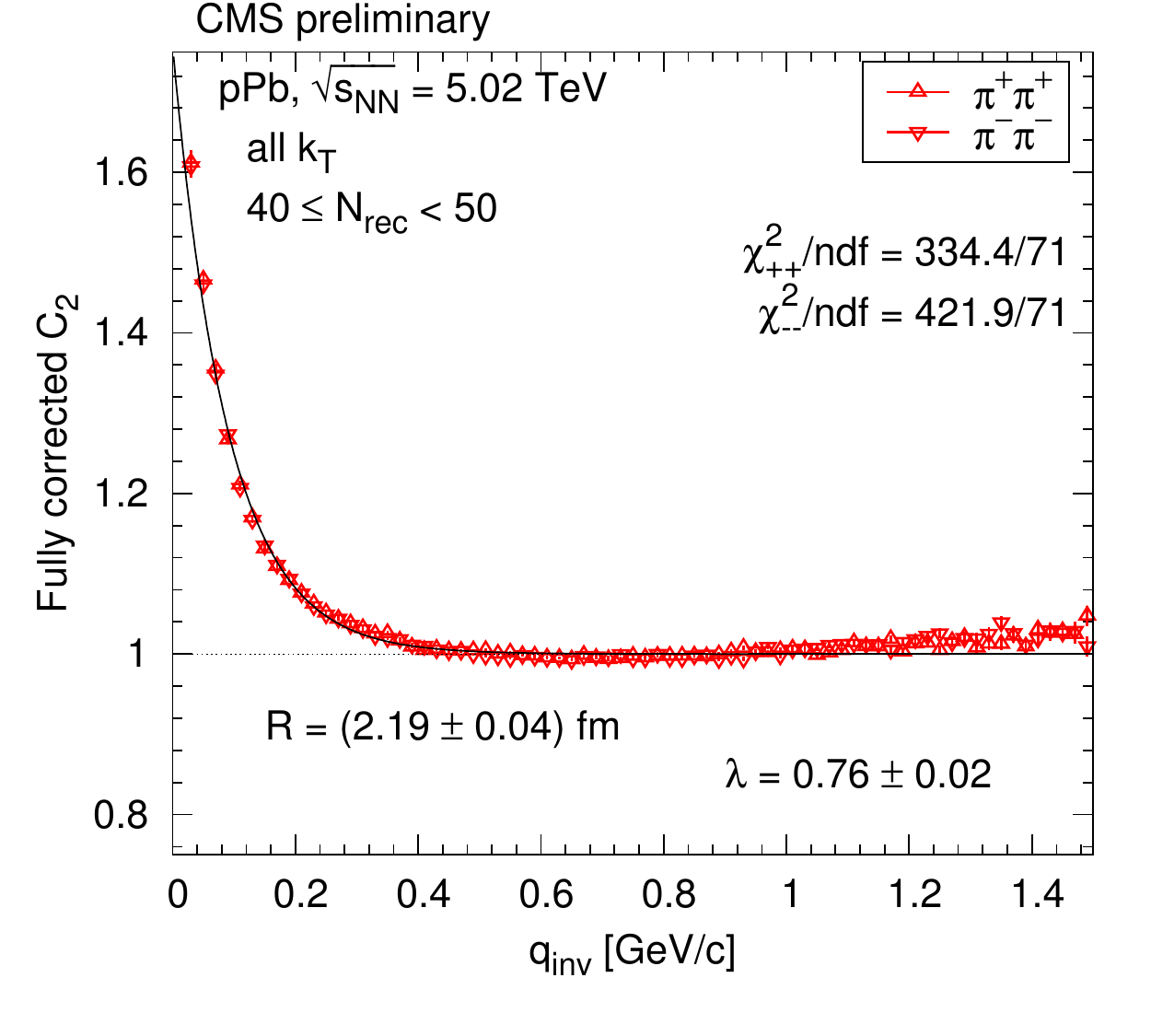}
  \includegraphics[width=0.49\textwidth]{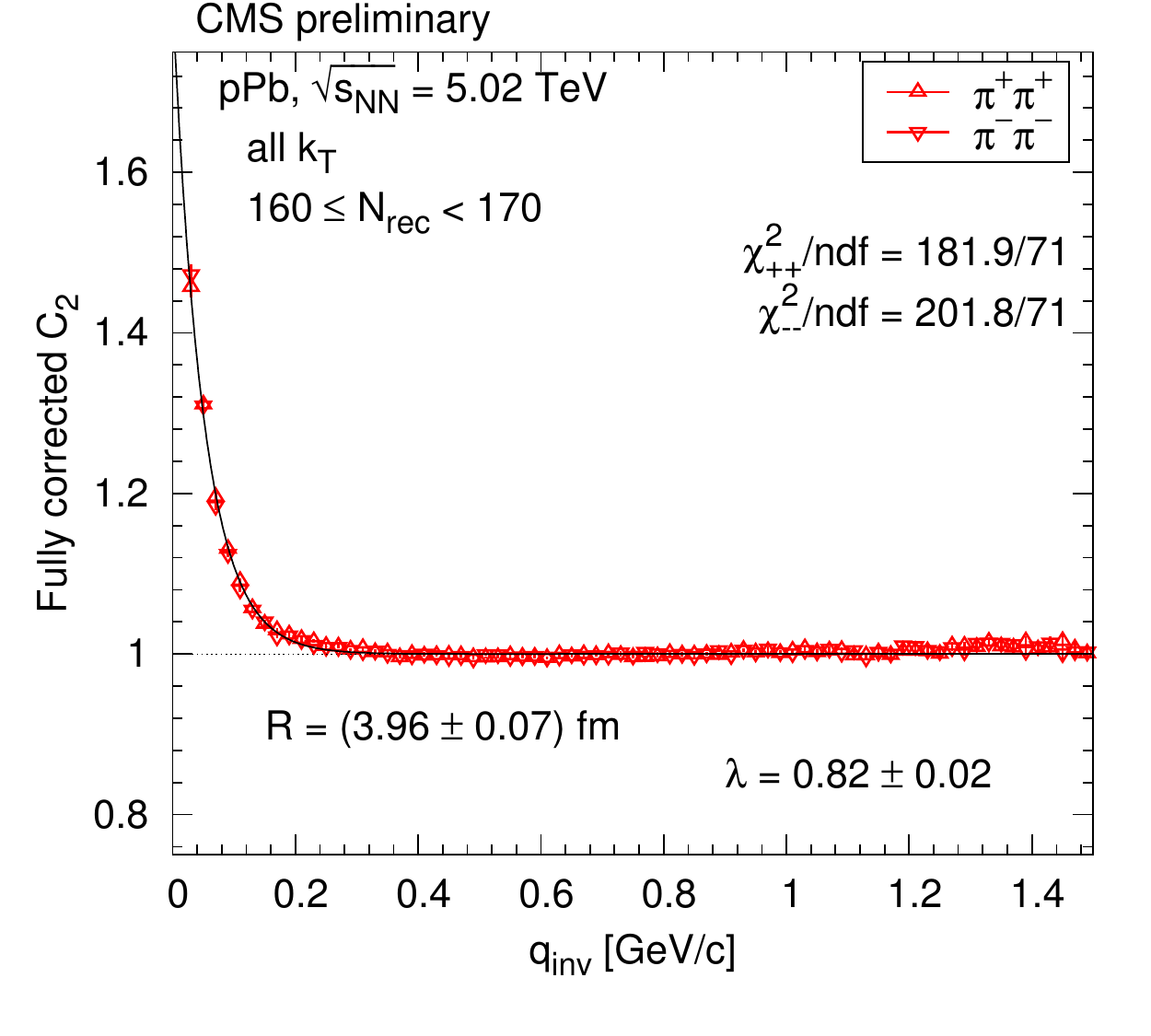}

  \includegraphics[width=0.49\textwidth]{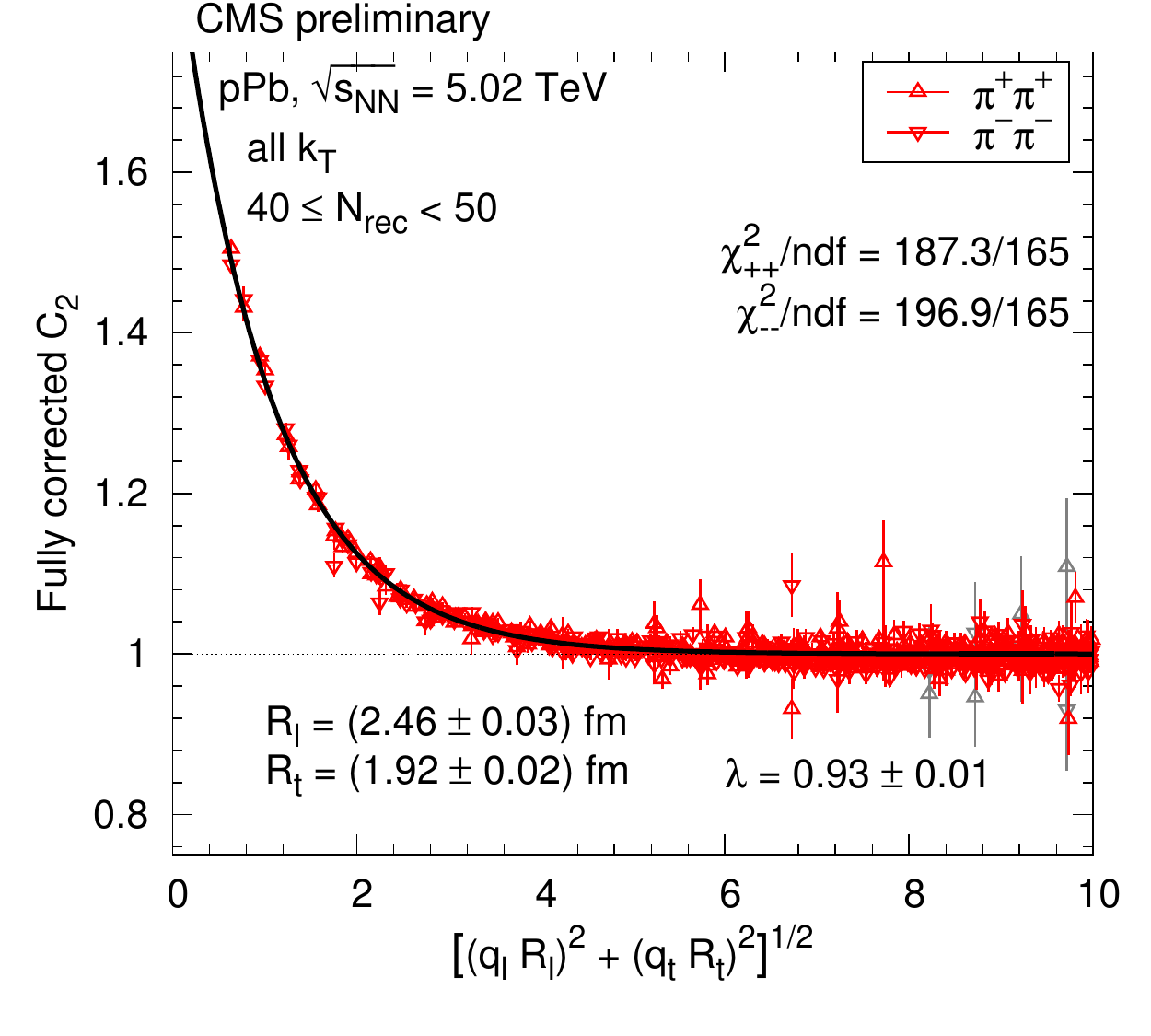}
  \includegraphics[width=0.49\textwidth]{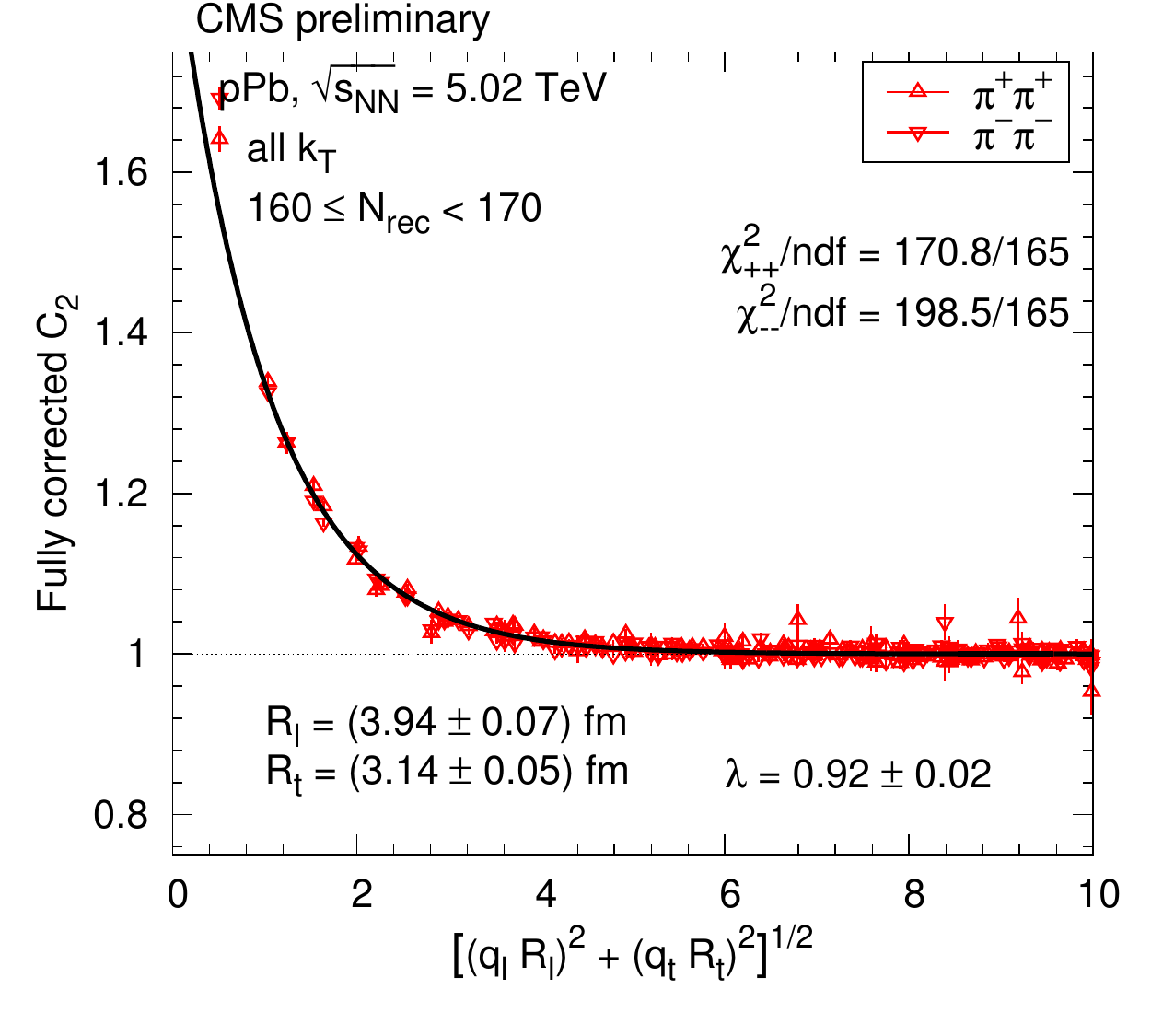}

  \includegraphics[width=0.49\textwidth]{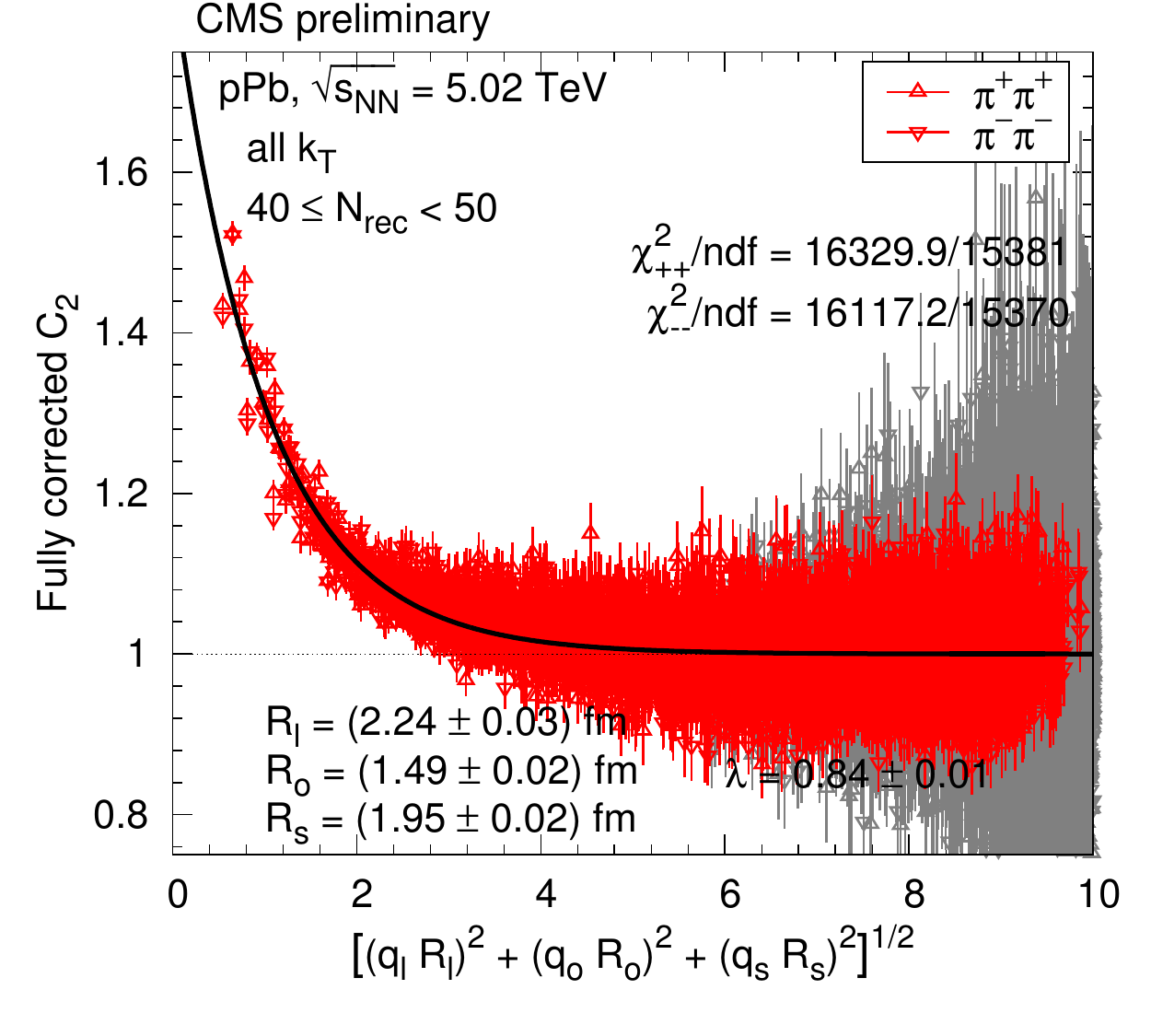}
  \includegraphics[width=0.49\textwidth]{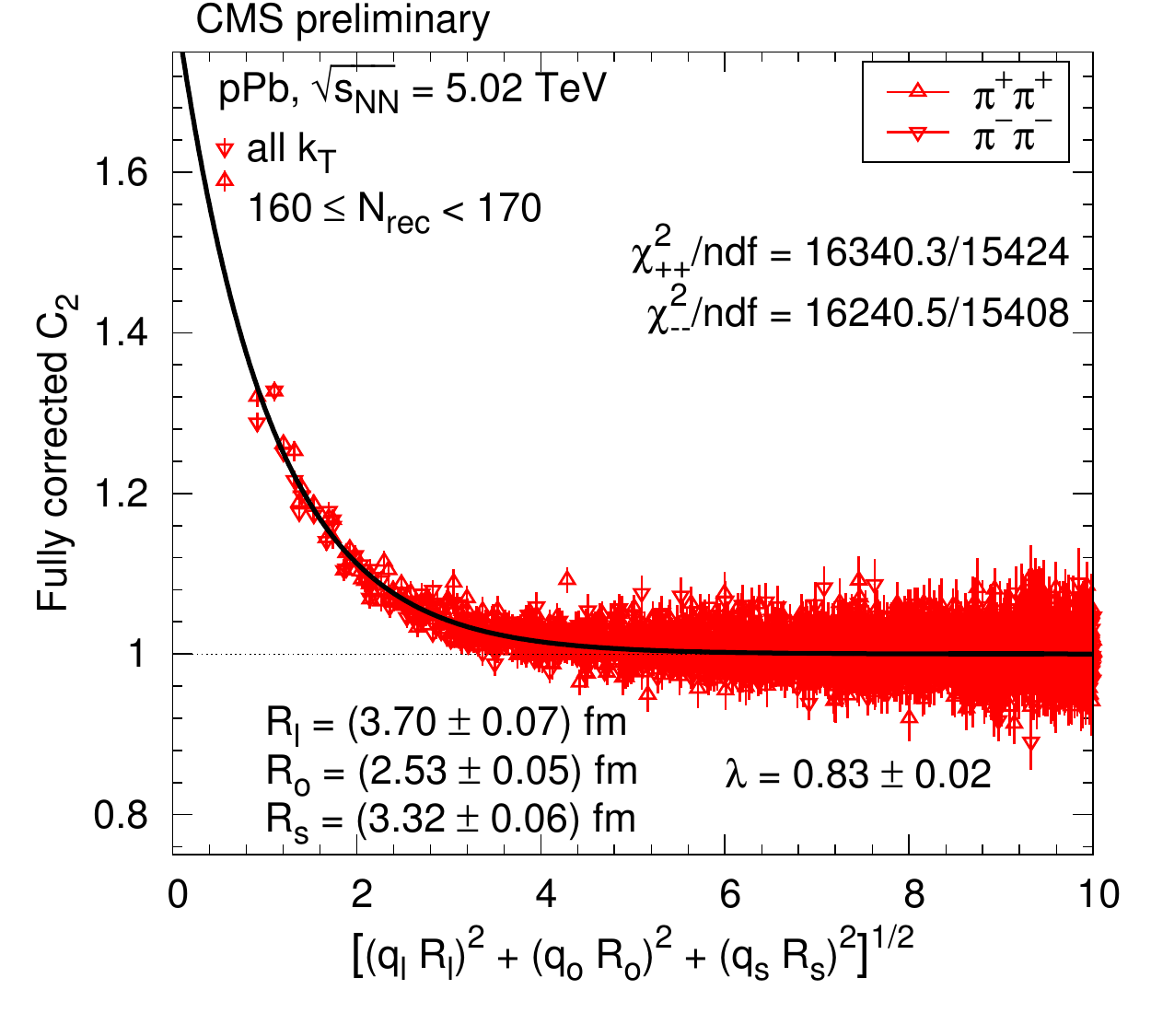}
 \end{center}

 \vspace{-0.2in}
 \caption{The like-sign correlation function of pions (red triangles) corrected
for Coulomb interaction and cluster contribution (mini-jets and multi-body
resonance decays) as a function of $\qi$ or the combined momentum, in some
selected $\Nr$ bins for all $\kt$~\cite{CMS:2014mla}. The solid curves indicate
fits with the exponential Bose-Einstein parametrization.}

 \label{fig:pi_pPb_5TeVc/qmix_}

\end{figure}

\begin{figure}

 \begin{center}
  \includegraphics[width=0.49\textwidth]{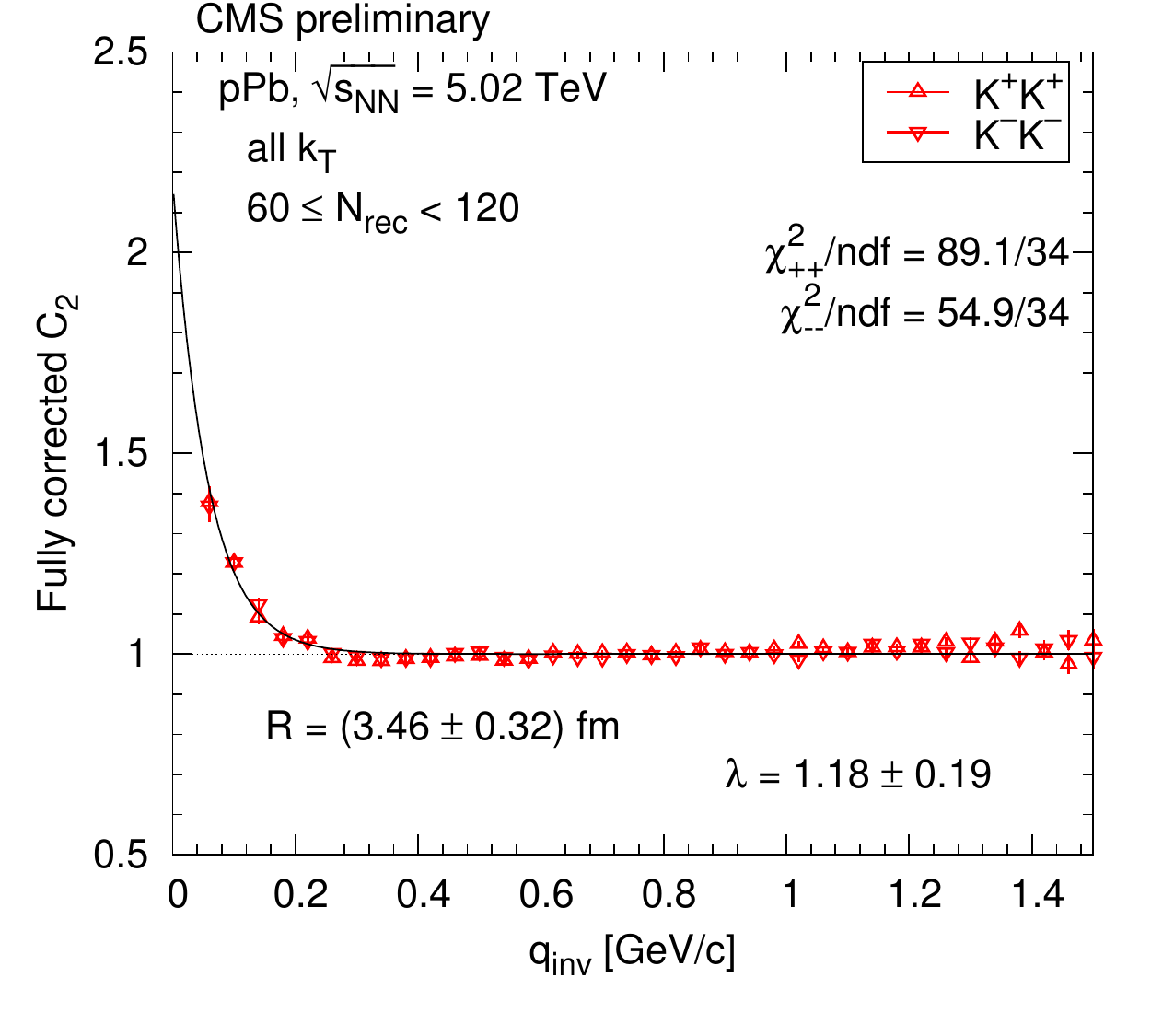}
  \includegraphics[width=0.49\textwidth]{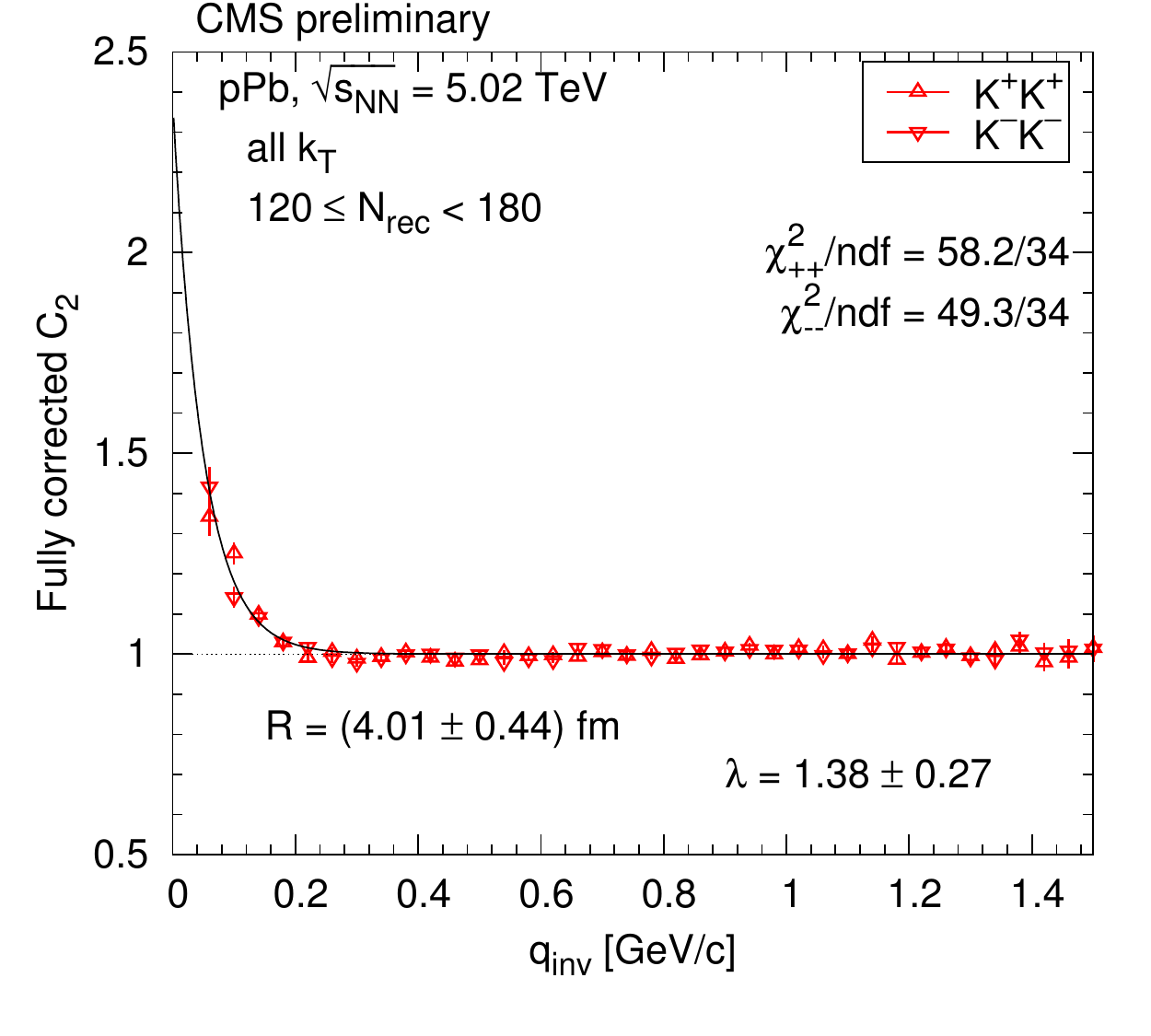}
 \end{center}

 \vspace{-0.2in}
 \caption{The like-sign correlation function of kaons (red triangles) corrected
for Coulomb interaction and cluster contribution (mini-jets and multi-body
resonance decays) as a function of $\qi$, in some selected $\Nr$ bins for all
$\kt$~\cite{CMS:2014mla}. The solid curves indicate fits with the Bose-Einstein
parametrization.}

 \label{fig:ka_pPb_5TeVc/qmix_}

\end{figure}

The cluster contribution can be also extracted in the case of a like-sign
correlation function, if the momentum scale of the Bose-Einstein correlation
and that of the cluster contribution ($\approx$\,0.4~\GeVc) are different enough.
An important element in both mini-jet and multi-body resonance decays is the
conservation of electric charge that results in a stronger correlation for
unlike-sign pairs than for like-sign pairs.
Hence the cluster contribution is expected to be also present for like-sign
pairs, with similar shape but a somewhat smaller amplitude.
The form of the cluster-related contribution obtained from unlike-sign pairs,
but now multiplied by the extracted relative amplitude $z$, is used to fit the
like-sign correlations.
A selection of correlation functions and fits are shown in
Figs.~\ref{fig:pi_pPb_5TeVc/qmix_} and \ref{fig:ka_pPb_5TeVc/qmix_}.

In the case of two and three dimensions the measured unlike-sign correlation
functions show that instead of $\qi$, the length of the weighted sum of
$\vec{q}$ components is a better common variable.

\newpage

\section{Results}

\label{sec:results}

The systematic uncertainties are dominated by two sources: the dependence of
the final results on the way the background distribution is constructed, and
the uncertainties of the amplitude $z$ of the cluster contribution for
like-sign pairs with respect to those for unlike-sign ones.

\begin{figure}

 \begin{center}
  \includegraphics[width=0.49\textwidth]{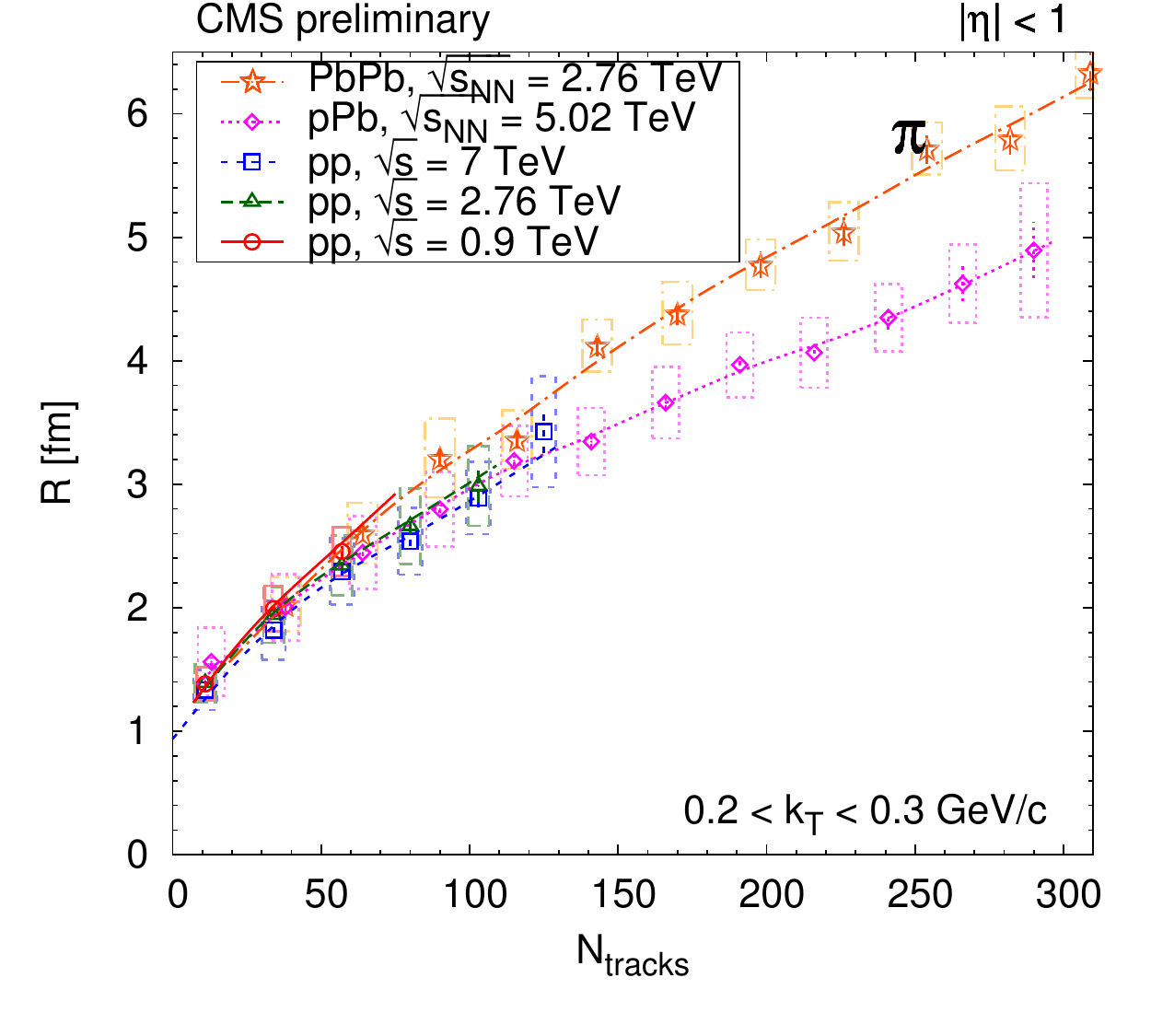}
  \includegraphics[width=0.49\textwidth]{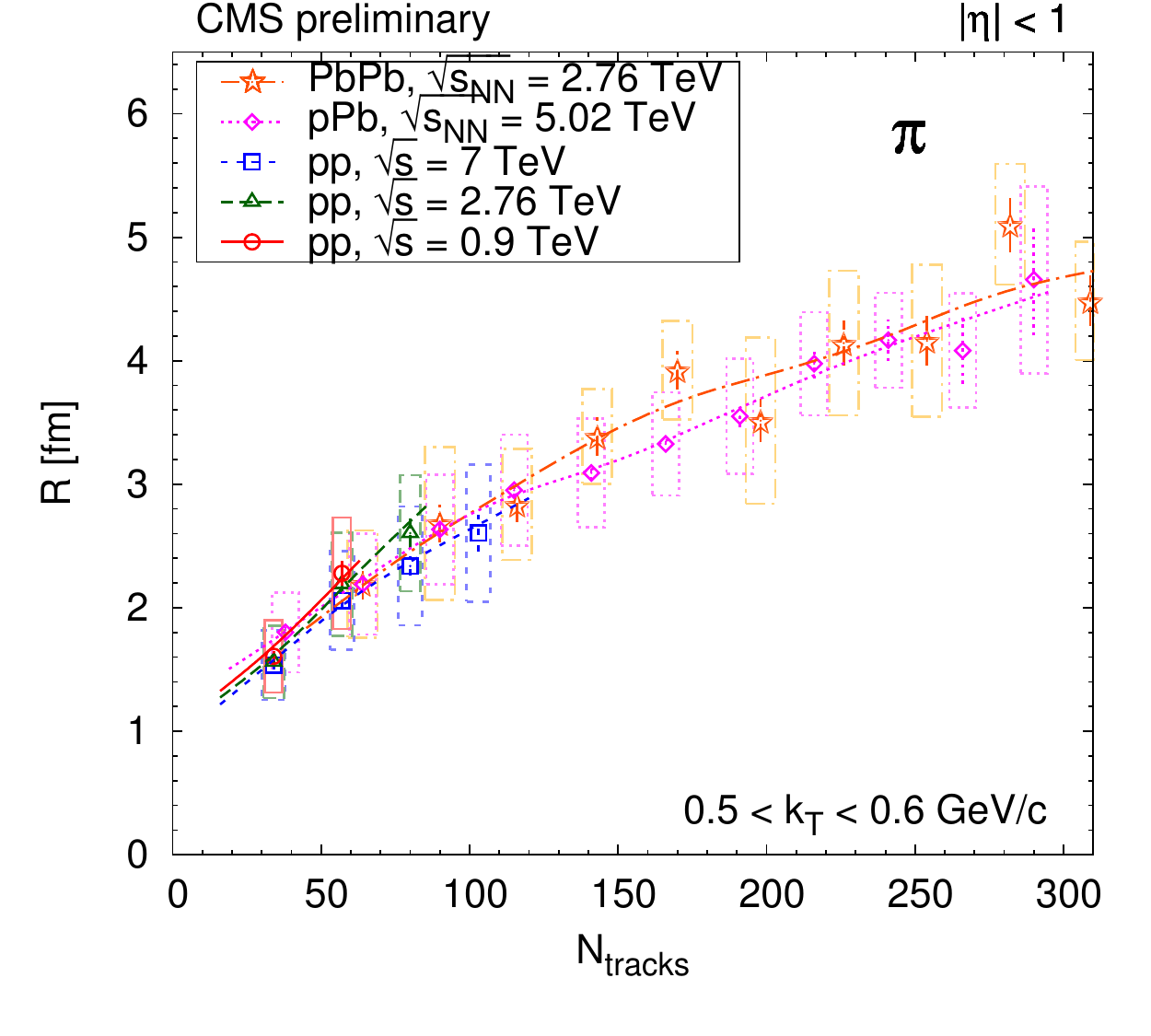}

  \includegraphics[width=0.49\textwidth]{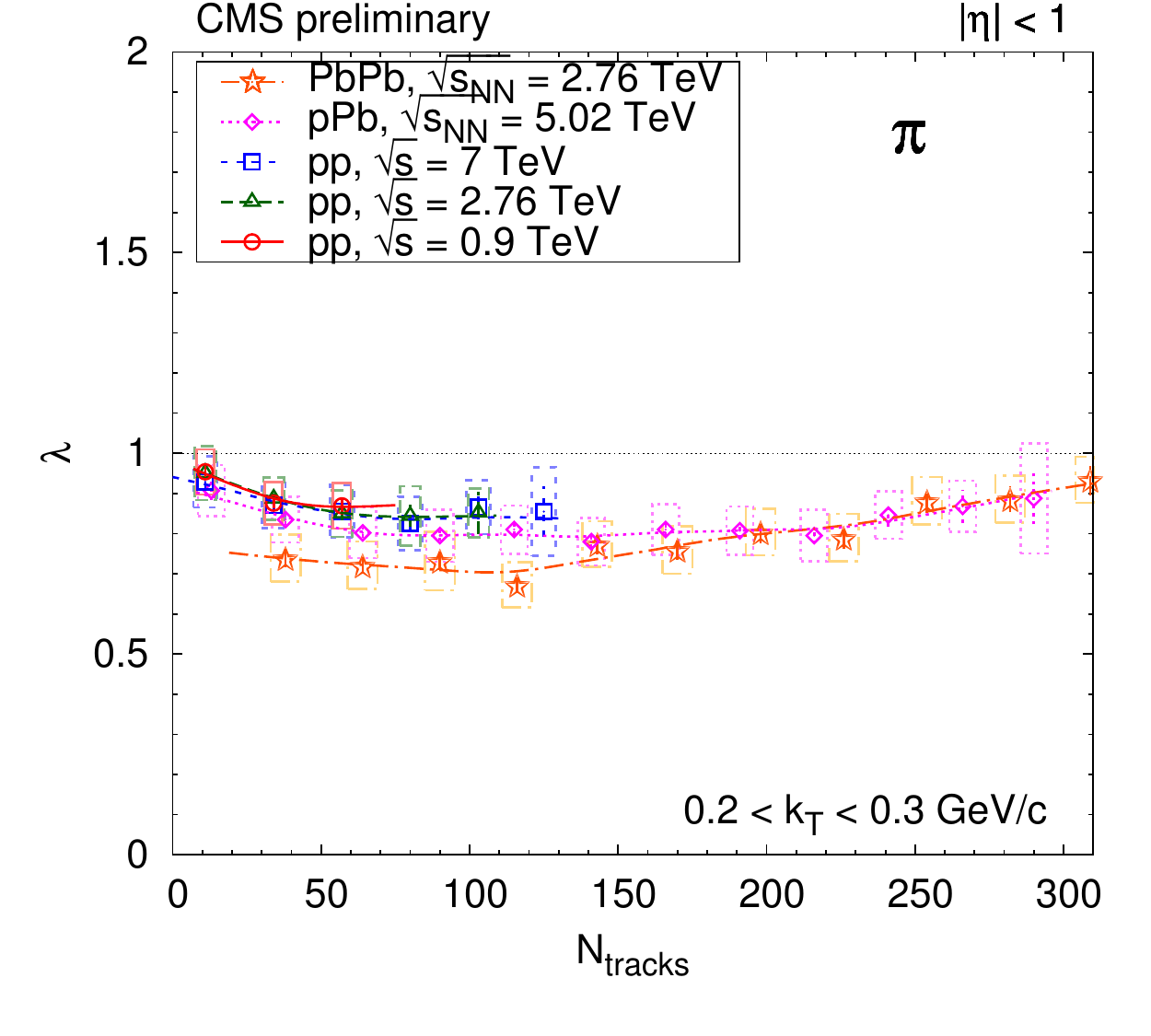}
  \includegraphics[width=0.49\textwidth]{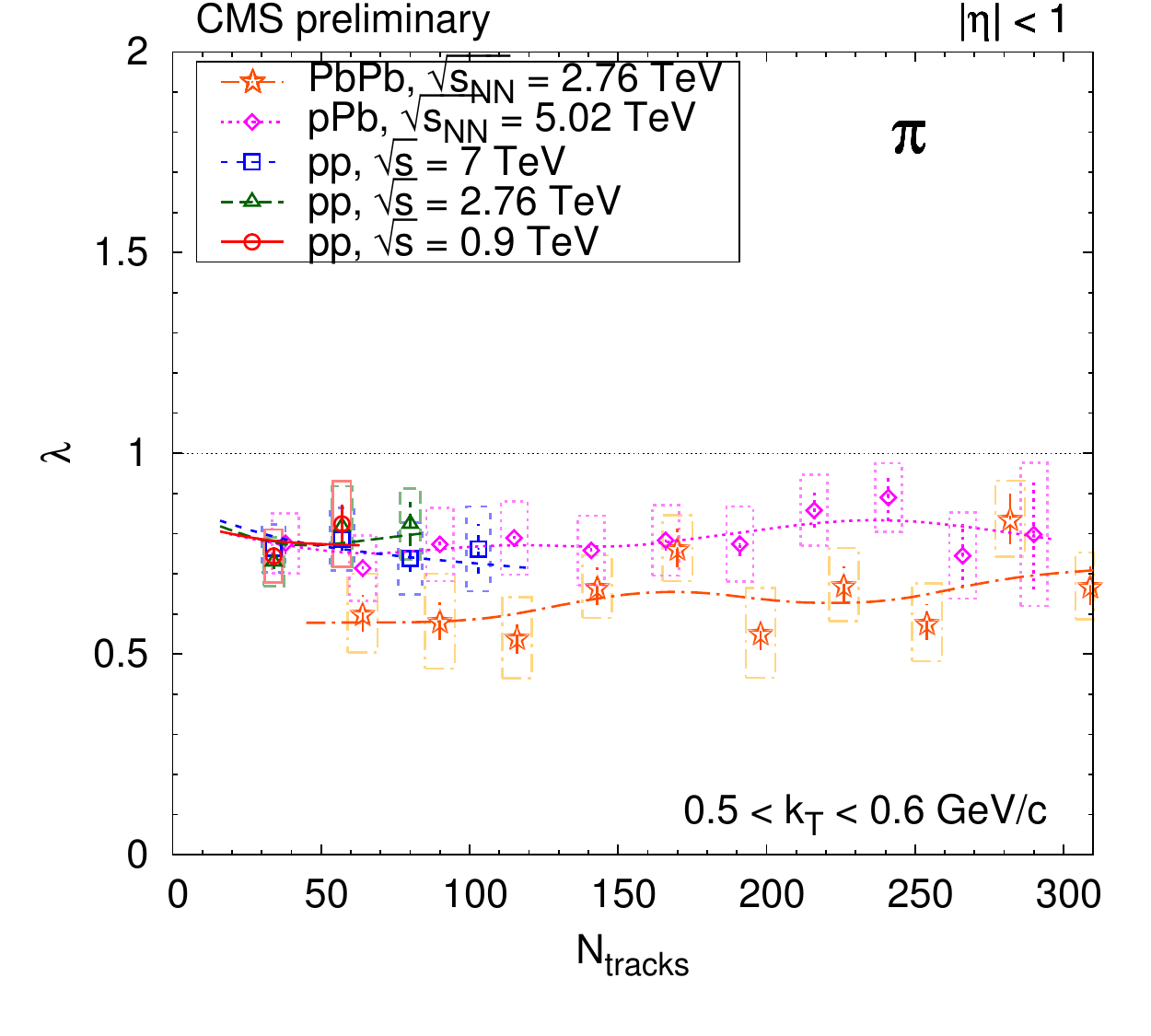}
 \end{center}

 \vspace{-0.2in}
 \caption{The $\Nt$ dependence of the one-dimensional pion radius (top) and the
one-dimensional pion chaoticity parameter (bottom), shown here for
several $\kt$ bins, for all studied reactions~\cite{CMS:2014mla}. \ldg}

 \label{fig:pi_radius_qinv}

\end{figure}

\begin{figure}

 \begin{center}
  \includegraphics[width=0.49\textwidth]{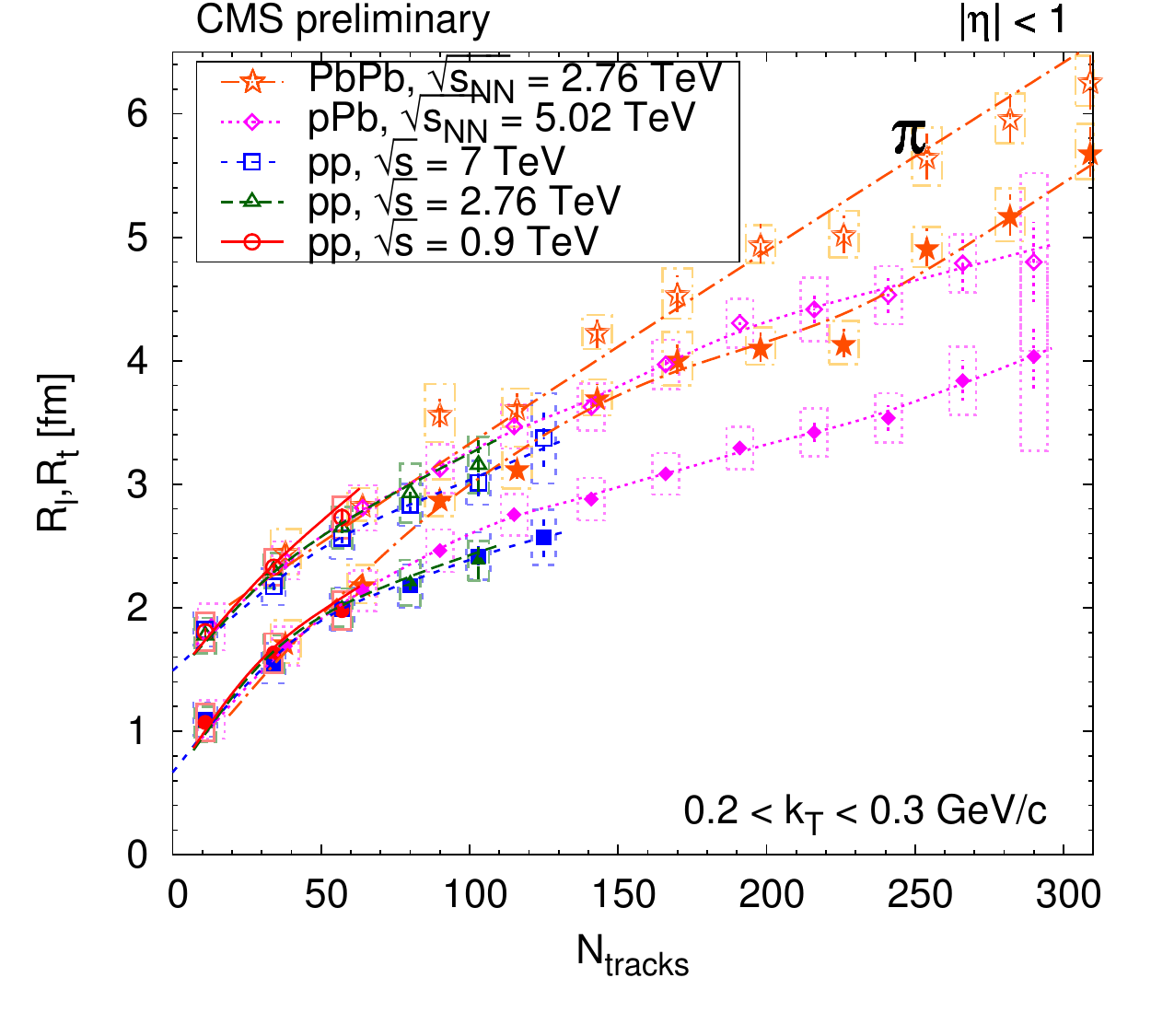}
  \includegraphics[width=0.49\textwidth]{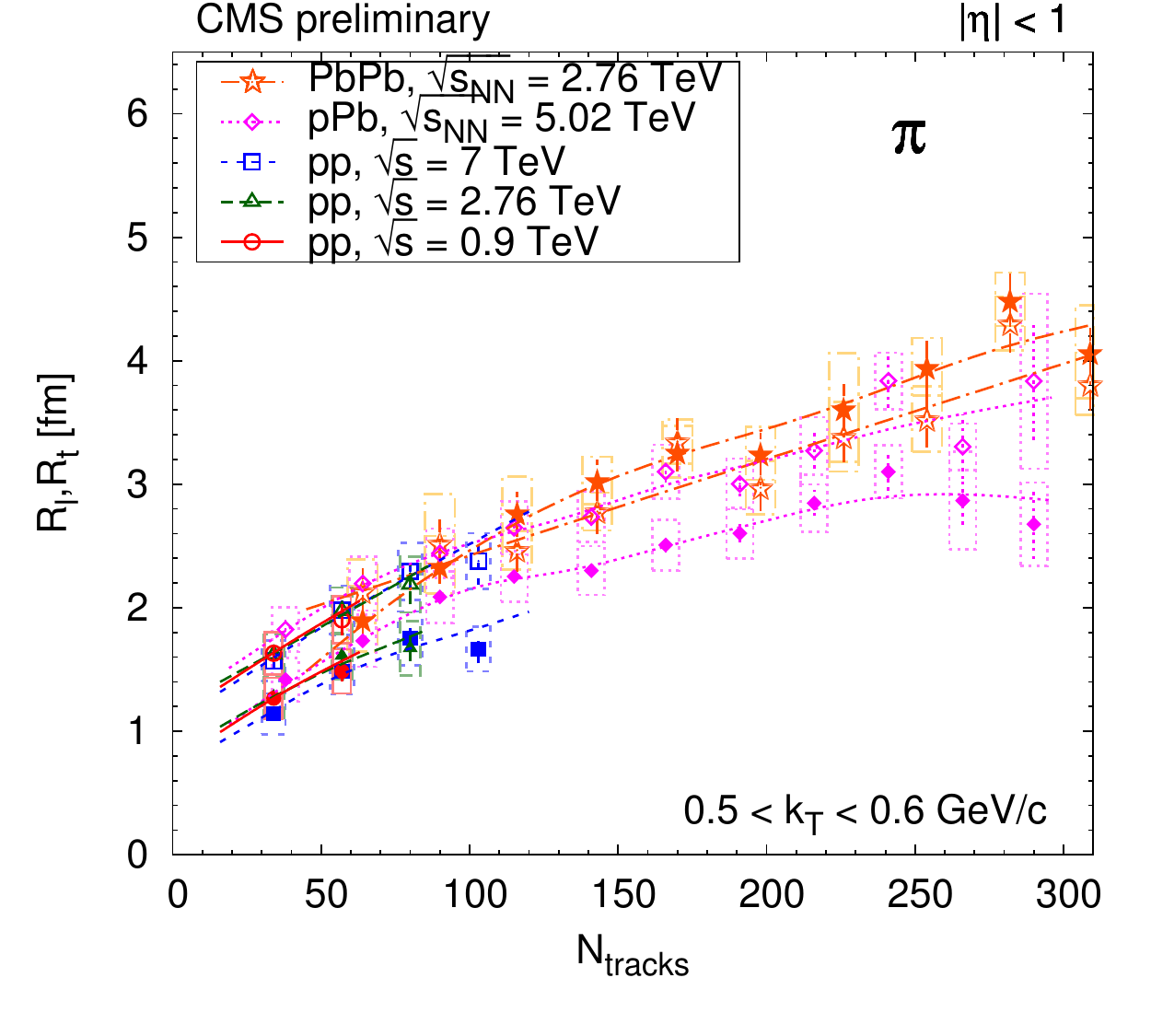}
 \end{center}

 \vspace{-0.2in}
 \caption{The $\Nt$ dependence of the two-dimensional pion radii ($R_l$ -- open
symbols, $R_t$ -- closed symbols), shown here for several $\kt$ bins, for all
studied reactions~\cite{CMS:2014mla}. \ldg}

 \label{fig:pi_radius_qlqt}

\end{figure}

The characteristics of the extracted one- and two-dimensional correlation
functions as a function of the transverse pair momentum $\kt$ and of the
charged-particle multiplicity $\Nt$ (in the range $|\eta| <$ 2.4 in the
laboratory frame) of the event are presented here. Three-dimensional results
are detailed in Ref.~\cite{CMS:2014mla}.
In all the following plots
(Figs.~\ref{fig:pi_radius_qinv}--\ref{fig:radius12_scale}), the results of
positively and negatively charged hadrons are averaged. For clarity, values and
uncertainties of the neighboring $\Nt$ bins were averaged two by two, and only
the averages are plotted. The central values of radii and chaoticity parameter
$\lambda$ are given by markers. The statistical uncertainties are indicated by
vertical error bars, the combined systematic uncertainties (choice of
background method; uncertainty of the relative amplitude $z$ of the cluster
contribution; low $q$ exclusion) are given by open boxes. Unless indicated, the
lines are drawn to guide the eye (cubic splines whose coefficients are found by
weighing the data points with the inverse of their squared statistical
uncertainty).

The extracted exponential radii for pions increase with increasing $\Nt$ for
all systems and center-of-mass energies studied, for one, two, and three
dimensions alike. Their values are in the range 1--5 fm, reaching highest
values for very high multiplicity pPb, also for similar multiplicity PbPb
collisions. The $\Nt$ dependence of $R_l$ and $R_t$ is similar for pp and pPb
in all $\kt$ bins, and that similarity also applies to peripheral PbPb if $\kt >
$ 0.4~\GeVc.  In general there is an ordering, $R_l > R_t$, and $R_l
> R_s > R_o$, thus the pp and pPb source is elongated in the beam direction. In
the case of peripheral PbPb the source is quite symmetric, and shows a slightly
different $\Nt$ dependence, with largest differences for $R_t$ and $R_o$, while
there is a good agreement for $R_l$ and $R_s$. The most visible divergence
between pp, pPb and PbPb is seen in $R_o$ that could point to the differing
lifetime of the created systems in those collisions.
 
The kaon radii also indicate some increase with $\Nt$ (not shown), although its
magnitude is smaller than that for pions. Longer lived resonances and
rescattering may play a role here.

\subsection{Scaling}

\begin{figure}

 \begin{center}
  \includegraphics[width=0.49\textwidth]{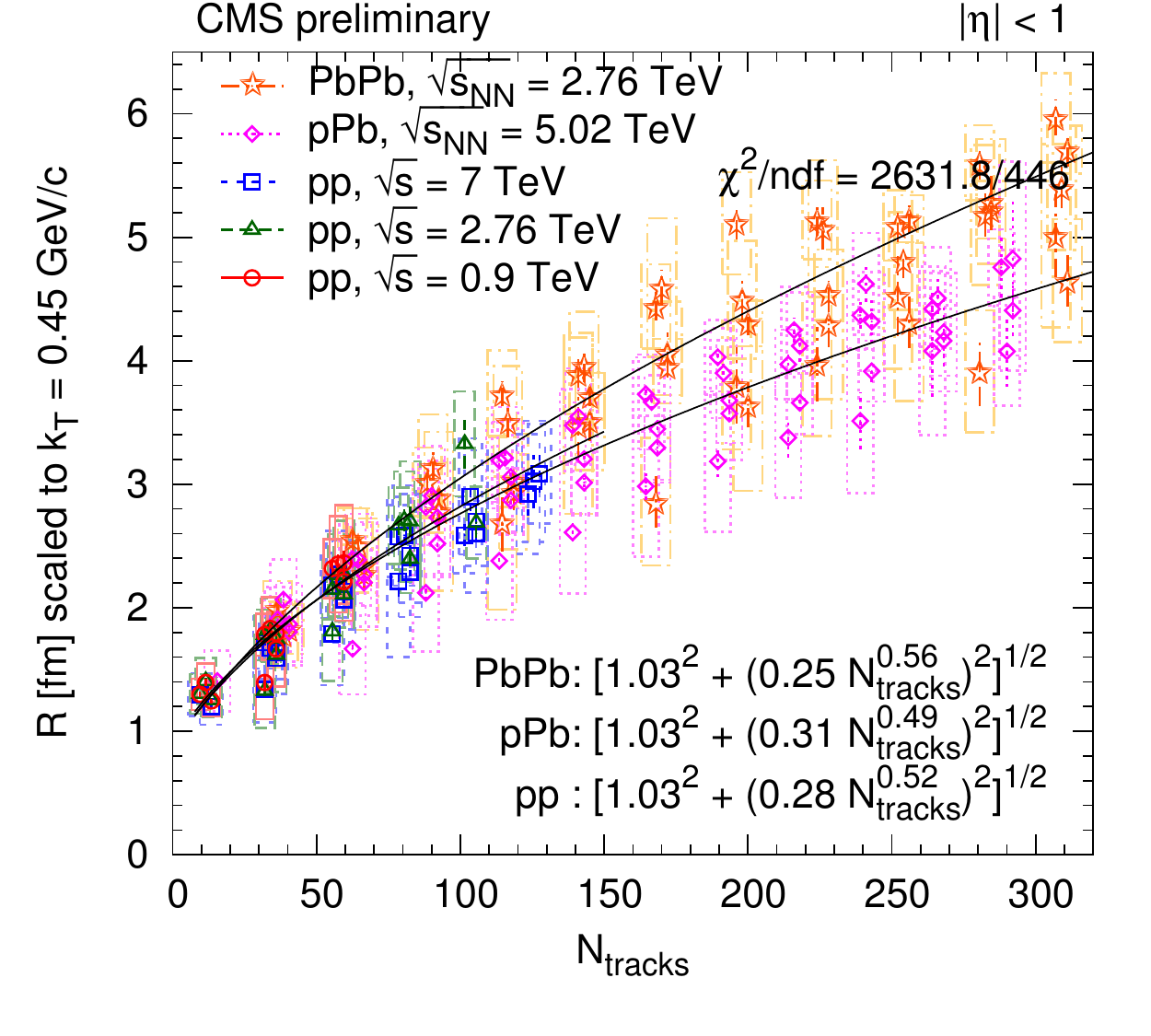}
  \includegraphics[width=0.49\textwidth]{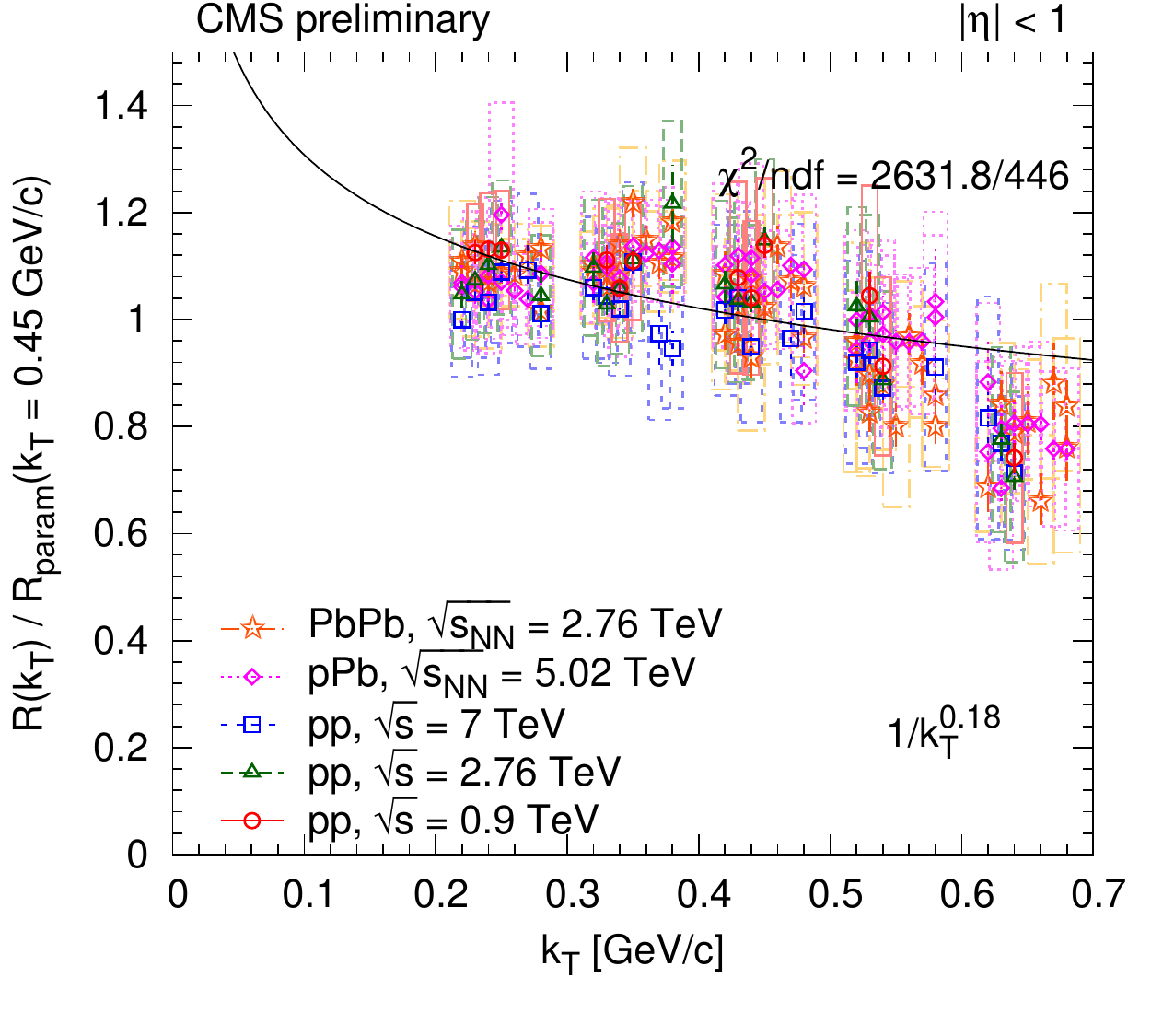}

  \includegraphics[width=0.49\textwidth]{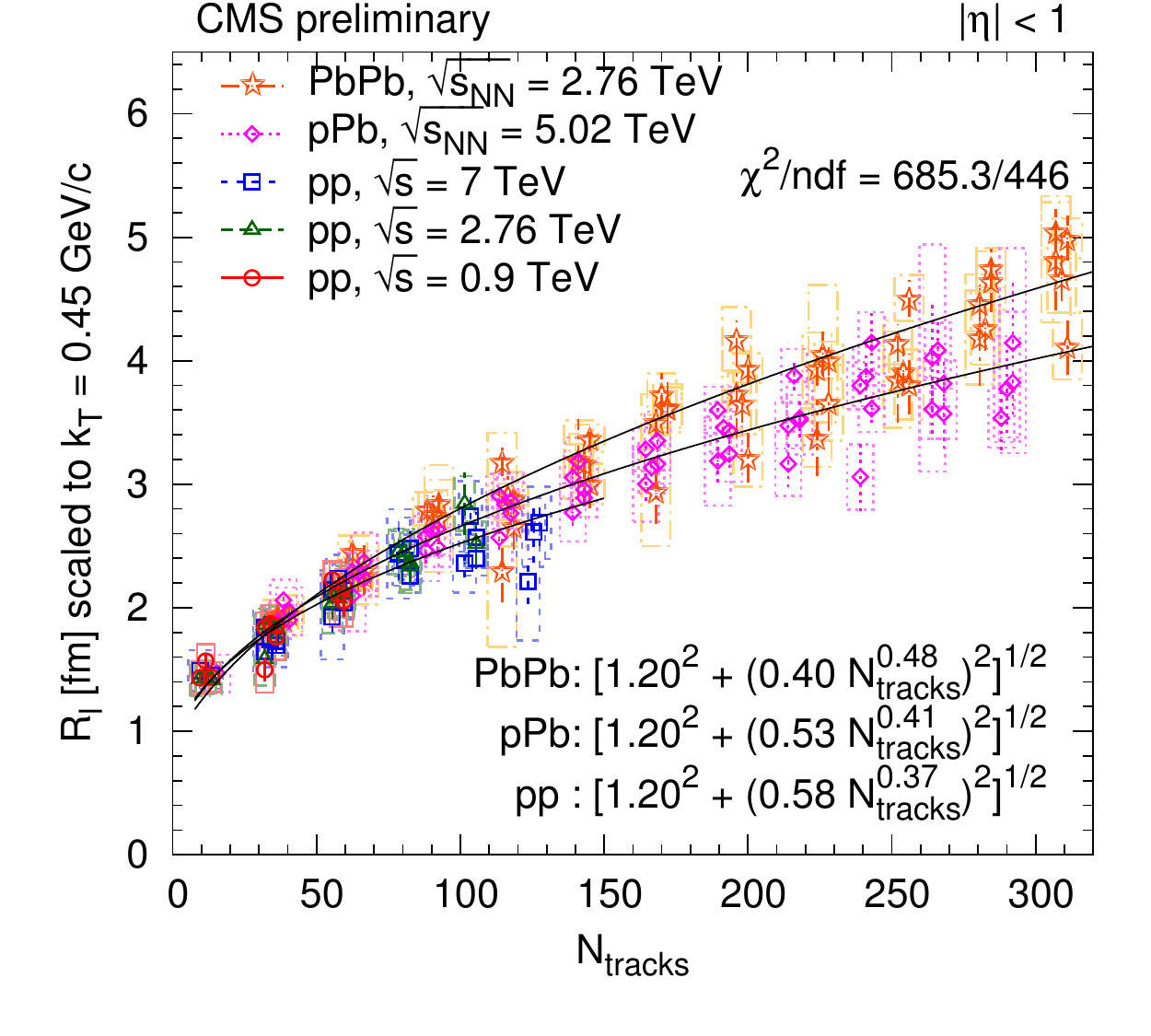}
  \includegraphics[width=0.49\textwidth]{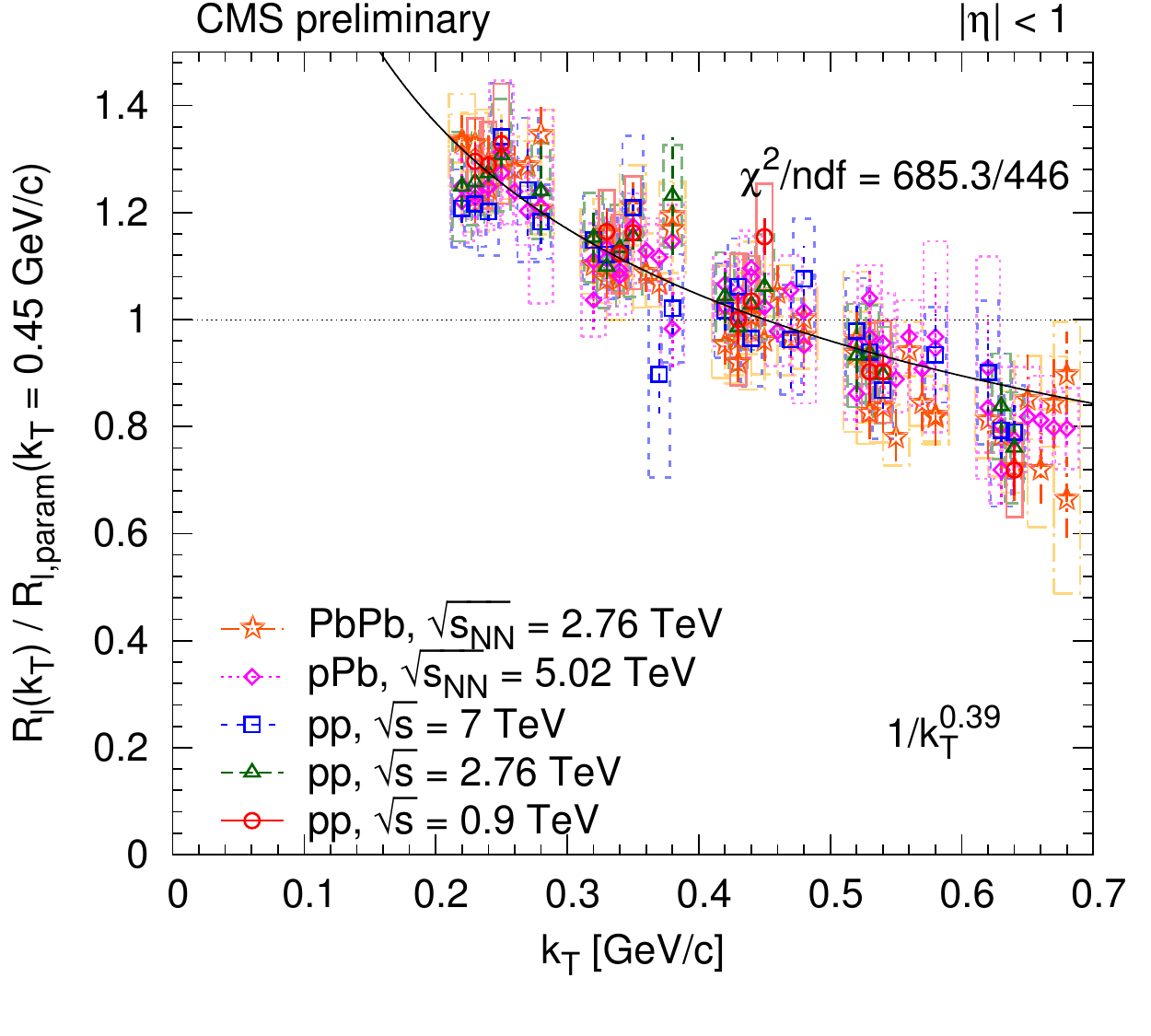}

  \includegraphics[width=0.49\textwidth]{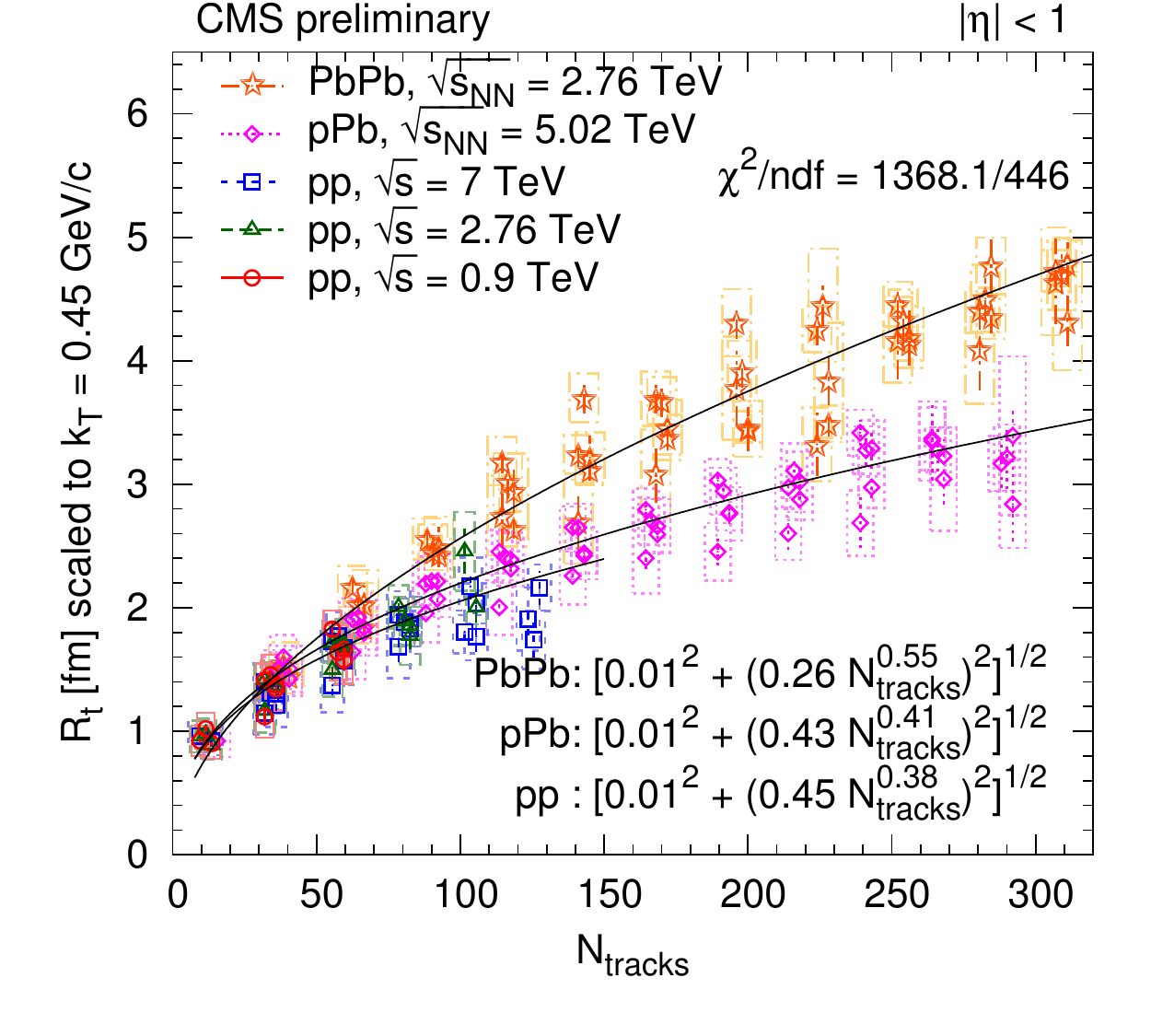}
  \includegraphics[width=0.49\textwidth]{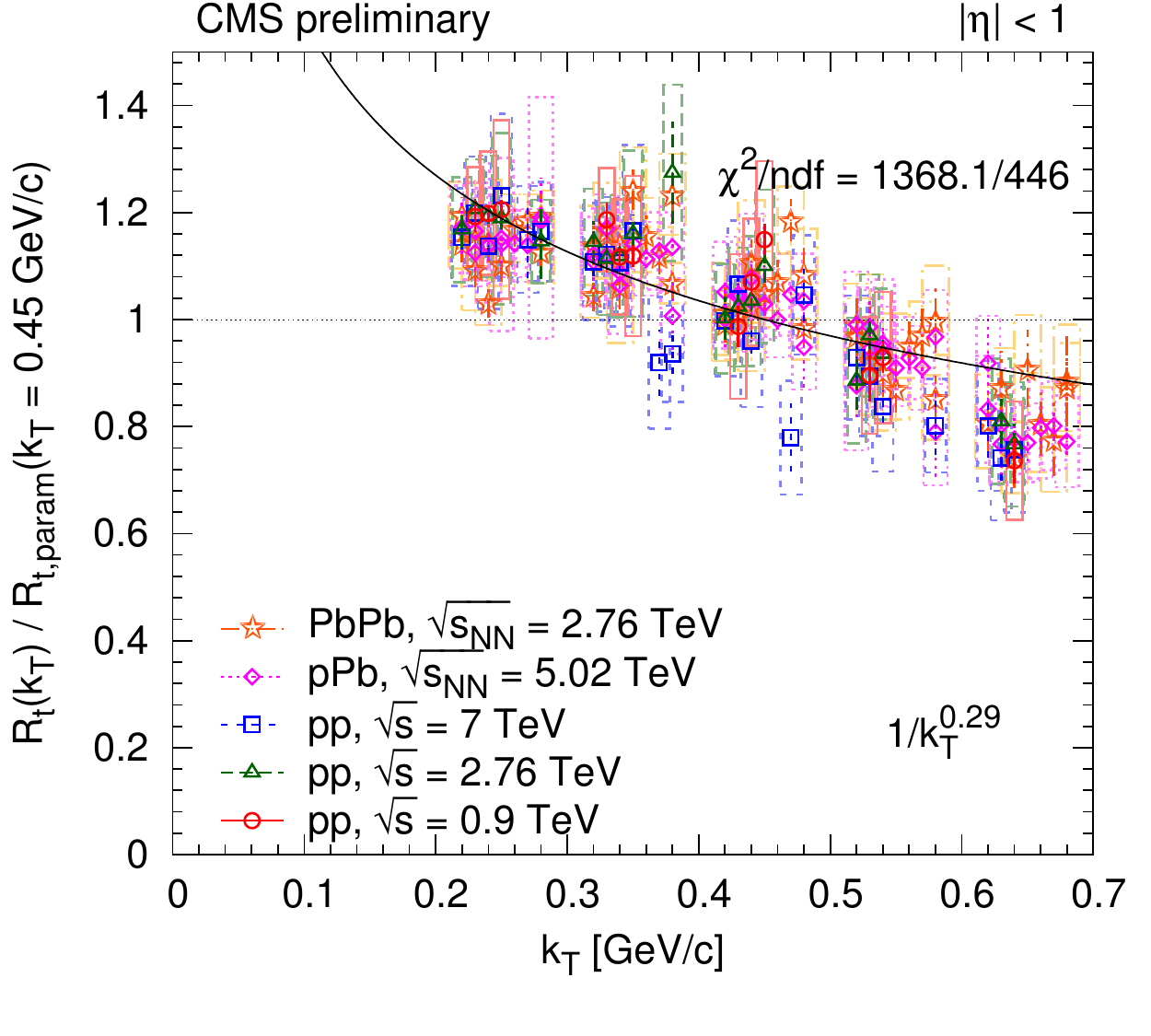}
 \end{center}

 \caption{Left: radius parameters as a function of $\Nt$ scaled to $\kt =
$ 0.45~\GeVc\ with help of the parametrization $R_\text{param}$
(Eq.~\eqref{eq:frakR}).  Right: ratio of the radius parameter and the value of
the parametrization $\Rp$ (Eq.~\eqref{eq:frakR}) at $\kt =$ 0.45~\GeVc\ as a
function of $\kt$.  (Points were shifted to left and to right with respect to
the center of the $\kt$ bin for better visibility.) Upper row: $R$ from the
one-dimensional ($\qi$) analysis.  Middle row: $R_l$ from the two-dimensional
$(\ql,\qt)$ analysis.  Bottom row: $R_t$ from the two-dimensional $(\ql,\qt)$
analysis. Fit results are indicated in the figures~\cite{CMS:2014mla}, for
details see text.}

 \label{fig:radius12_scale}

\end{figure}

The extracted radii are in the range 1--5 fm, reaching highest values for very
high multiplicity pPb, also for similar multiplicity PbPb collisions, and
decrease with increasing $\kt$. By fitting the radii with a product of two
independent functions of $\Nt$ and $\kt$, the dependences on multiplicity and
pair momentum appear to factorize. In some cases the radii are less
sensitive to the type of the colliding system and center-of-mass energy.
Radius parameters as a function of $\Nt$ at $\kt =$ 0.45~\GeVc\ are shown in the
left column of Fig.~\ref{fig:radius12_scale}.
We have also fitted and plotted the following $\Rp$ functions

\begin{equation}
 \label{eq:frakR}
 {\Rp}(\Nt,\kt) = 
   \bigl[a^2 + (b \Nt^\beta)^2\bigr]^{1/2} \cdot
    \left(0.2~\mathrm{GeV}/c/\kt\right)^\gamma,
\end{equation}

\noindent where the minimal radius $a$ and the exponents
$\gamma$ of $\kt$ are kept the same for a given radius component, for all
collision types.
This choice of parametrization is based on previous results~\cite{Lisa:2005js}.
The minimal radius can be connected to the size of the proton, while the
power-law dependence on $\Nt$ is often attributed to the freeze-out density of
hadrons.
The ratio of radius parameter and the value of the above parametrization at
$\kt =$ 0.45~\GeVc\ as a function $\kt$ is shown in the right column of
Fig.~\ref{fig:radius12_scale}.

\section{Conclusions}

The similarities observed in the $\Nt$ dependence may point to a common
critical hadron density in pp, pPb, and peripheral PbPb collisions, since the
present correlation technique measures the characteristic size of the system
near the time of the last interactions.

\vspace{6pt} 

\acknowledgments{This work was supported by the Hungarian Scientific Research
Fund (K~109703), and the Swiss National Science Foundation (SCOPES 152601).}

\externalbibliography{yes}
\bibliography{sikler_cms_bgl17}

\end{document}